\documentclass[iop]{emulateapj}

\usepackage{apjfonts}
\usepackage{pbox}
\usepackage{bm}
\usepackage{CJK}
\usepackage[breaklinks,colorlinks,citecolor=blue,linkcolor=magenta]{hyperref}
\usepackage{soul}

\newcommand{\set}[1]{\bm{#1}}
\newcommand{\starlabel}{\ell}
\newcommand{\labels}{\set{\starlabel}}

\shorttitle{$>20$ Element Abundances with Low-Resolution Spectra}
\shortauthors{Ting et al.}

\begin{document}

\begin{CJK*}{UTF8}{gbsn}
\title{Prospects for Measuring Abundances of $>20$ Elements with \\ Low-resolution Stellar Spectra}
\author{Yuan-Sen Ting (丁源森)\altaffilmark{1,2}, Charlie Conroy\altaffilmark{1}, Hans-Walter Rix\altaffilmark{3}, Phillip Cargile\altaffilmark{1}}
\altaffiltext{1}{Harvard--Smithsonian Center for Astrophysics, 60 Garden Street, Cambridge, MA 02138, USA}
\altaffiltext{2}{Research School of Astronomy $\&$ Astrophysics, The Australian National University, Cotter Road, Weston Creek, ACT 2611, Australia}
\altaffiltext{3}{Max Planck Institute for Astronomy, K\"onigstuhl 17, D-69117 Heidelberg, Germany}

\slugcomment{Submitted to ApJ}

%
%
%
%
%
%
\begin{abstract}
Understanding the evolution of the Milky Way calls for the precise abundance determination of many elements in many stars. A common perception is that deriving more than a few elemental abundances ([Fe/H], [$\alpha$/Fe], perhaps [C/H], [N/H]) requires medium-to-high spectral resolution, $R \gtrsim \,$10$,$000, mostly to overcome the effects of line blending. In recent work \citep{rix16,tin16} we presented an efficient and practical way to model the full stellar spectrum, even when fitting a large number of stellar labels simultaneously. In this paper we quantify to what precision the abundances of many different elements can be recovered, as a function of spectroscopic resolution and wavelength range. In the limit of perfect spectral models and spectral normalization, we show that the precision of elemental abundances is nearly independent of resolution, for a fixed exposure time and number of detector pixels; low-resolution spectra simply afford much higher S/N per pixel and generally larger wavelength range in a single setting. We also show that estimates of most stellar labels are not strongly correlated with one another once $R \gtrsim \,$1$,$000. Modest errors in the line spread function, as well as small radial velocity errors, do not affect these conclusions, and data driven models indicate that spectral (continuum) normalization can be achieved well enough in practice. These results, to be confirmed with an analysis of observed low-resolution data, open up new possibilities for the design of large spectroscopic stellar surveys and for the re-analysis of archival low-resolution datasets.
\end{abstract}
 
\keywords{methods: data analysis --- stars: abundances --- stars: atmospheres --- techniques: spectroscopic}

%
%
%
%
%
%

 \section{Introduction}
\label{sec:introduction}

Massively multiplexed stellar spectroscopic surveys are a central part of the current astronomy landscape, aimed at understanding stellar physics, the genesis of elements in the cosmos and the chemical/dynamical evolution of the Milky Way. This field is currently undergoing a revolution in the quality and quantity of spectra \citep[e.g., see review from][]{rix13}: current surveys aim to collect high quality spectra for millions of stars. But these extensive datasets bring new analysis and modeling challenges. Novel approaches are emerging \citep[e.g.,][]{nes15a,cas16,rix16,tin16} for turning these massive datasets into precise stellar labels, encompassing stellar parameters and elemental abundances of stars.

Spectral resolution, $R$, is a key parameter characterizing spectroscopic surveys, and the goal of this paper is to determine the resolution required to measure stellar labels to a desired precision. Traditionally, stellar spectroscopy has parsed itself into three resolution regimes: low-resolution with $R \lesssim \ $10$,$000, high-resolution with 10$,$000$ \, \lesssim R \lesssim \, $50$,$000, and ultra high-resolution with $R \gtrsim $50$,$000. Low-resolution $R \simeq \,$2$,$000--10$,$000 spectroscopic surveys, such as SEGUE \citep{yan09}, RAVE \citep{ste06}, Gaia Radial Velocity Spectrometer (RVS) \citep{rec16}, and LAMOST \citep{luo15}, have aimed at deriving fundamental stellar parameters such as $T_{\rm eff}$, $\log g$ and radial velocity. However, because essentially all stellar spectral lines are blended at low-resolution, only measurements of [Fe/H] and a few elements such as $[\alpha/{\rm Fe}]$, [C/H], [N/H] have been attempted systematically \citep[e.g.,][]{kir10,lee11,lee13,yan13,fra16}. But even with a limited number of stellar labels, these surveys are crucial because they can amass the statistical samples necessary to provide a global view of the Galaxy. For example, metallicity distribution functions can be used to infer star formation histories \citep[e.g.,][]{cas11,hay15}; metallicity gradients provide a window into the global dynamical history of stars \citep[e.g.,][]{sch09,gra15,kaw16} and the inside-out formation of the Milky Way \citep[e.g.,][]{min14,sch16}, and stellar age-metallicity-kinematic dispersion relations identify the extent to which stars become kinematically dispersed over time due to dynamical heating \citep[e.g.,][]{mar14,aum16,gra16}. 

High-resolution spectroscopic surveys such as the on-going APOGEE \citep{maj15}, GALAH \citep{des15} and Gaia-ESO \citep{smi14} surveys are collecting stellar spectra with $R \simeq \,$24$,$000. These surveys are designed to overcome the perceived shortcomings of their low-resolution counterparts, and aim to measure detailed elemental abundances of $10-40$ elements. Precise abundances for many elements are a key to understanding the chemical evolution of the Milky Way, as well as stellar nucleosynthesis. For example, core-collapse supernovae from massive stars produce relative overabundances of $\alpha$-capture elements \citep[e.g.,][]{woo95,lim03}, whereas type Ia supernovae produce overabundances of iron-peak elements \citep[e.g.,][also review from \citealt{nom13}]{iwa99}. Mass loss from AGB stars adds additional complexity to the chemical evolution of the Milky Way \citep[e.g.,][]{kar14,ven15}. \citet{tin12a} used principal component analysis and demonstrated there are at least seven pathways for galaxies to collect their metals. One goal of deriving multi-elemental abundances for many stars is to unravel the contributions from these different channels at various evolutionary stages of the Milky Way. 

Ultra high-resolution spectra, with $R \gtrsim \,$50$,$000, are the gold standard for measuring precise and accurate stellar parameters and detailed abundances. At this resolution many of the strong stellar absorption lines are unblended. However, such spectra make high demands on instrumentation and exposure time. For this reason high-resolution surveys \citep[e.g.,][]{fis05,ben14,jof14,jof15,bre15,hei15} contain far fewer stars than medium and low-resolution surveys.

An exciting application of precise abundance measurements for many elements is the concept of chemical tagging \citep{fre02}: if stars born from the same molecular cloud share the same or very similar elemental abundances \citep[as suggested by recent observational works][]{bov15,liu16}, then elemental abundances can serve as a birth-tag for each star. The goal of ``strong'' chemical tagging is to look for stellar siblings by searching for similarities in chemical space \citep{fre02}. ``Strong'' chemical tagging has proven to be challenging and is yet to be realized in practice, in part because it requires a vast sample size and very precise elemental abundances \citep{lin13,tin15a}. But a weaker form of chemical tagging has been demonstrated to be viable \citep[e.g.,][]{qui15,hog16,mar16}. ``Weak'' chemical tagging uses precise measurement of elemental abundances to separate various groups of stars. For example, dwarf galaxies in the Milky Way are separable both from each other and from the Milky Way stellar halo in [$\alpha$/Fe]-[Fe/H] space \citep[e.g.,][]{ven04}; globular cluster stars have unique abundance patterns that allow their identification in the Milky Way bulge and stellar halo \citep[e.g.,][]{mar10,schi16}; and the thick disk, thin disk and halo stars can be well separated with their $\alpha$-enhancement measurements \citep[e.g.,][]{haw15,hay15}.

With strong scientific motivation for precisely measured elemental abundances of many ($\gtrsim20$) elements in many ($N>10^6$) stars, it is worth revisiting the optimal survey configuration to achieve these goals. In this paper, we will demonstrate that -- at a given exposure time or survey speed, and for a fixed number of detector pixels, low-resolution spectra can constrain comparably many elements and at the same precision as high-resolution spectra. Moreover, the estimates of elemental abundances show little correlation, once $R \gtrsim \,$1$,$000, even though the spectral lines are severely blended at low-resolution. These conclusions apply in the limit of very high quality spectral models, although the influence of bad pixels is smaller than often assumed. These results suggest new strategies for designing future generations of spectroscopic surveys.

This paper is structured as follows -- in Section~\ref{sec:basic-ideas}, we provide an overview of basic concepts and intuition related to spectra as a function of resolution, and describe how to model spectra in high dimensional space. We explore the information content of low-resolution spectra in Section~\ref{sec:theoretical-study}, and in Section~\ref{sec:real-study} we perform spectral fitting on synthetic spectral data with characteristics similar to the APOGEE survey. In Section~\ref{sec:discussions}, we discuss some implications of these results and highlight several caveats. We conclude in Section~\ref{sec:conclusions}.

%
%
%
%
%
%

\vspace{1cm}
\section{Basic ideas}
\label{sec:basic-ideas}

In this section we present the basic arguments for why there need not be significant loss of abundance information when choosing a spectroscopic survey with $R \sim \,$1$,$000, instead of $R \sim \,$100$,$000. The arguments presented in this section turn out to be fairly insensitive to the detailed wavelength range chosen for any spectroscopic survey. An important caveat throughout this work is the assumption of highly accurate spectral models. All commonly used {\it ab initio} stellar spectral models \citep[e.g.,][]{kur96,kur03,kur05,hau99,gus08} have important limitations, e.g., arising from incomplete and/or inaccurate atomic and molecular line parameters and the assumptions of 1D LTE \citep[see also][]{smi14}.  Therefore some of the results we present speak at present to the {\it information content}, in principle, of low-resolution spectra, but may require data-driven models \citep{nes15a,cas16} and/or improved {\it ab initio} spectral models in order to apply these results to real data. Moreover, in future work we will directly test many of our conclusions by fitting models to observations of stars with spectra obtained with a variety of spectral resolutions and wavelength ranges.

%
%
%
%
%
%

\subsection{The advantage of high-resolution spectra}
\label{sec:why-high-res}

Photospheric elemental abundances are encoded in the strengths of atomic and molecular absorption lines. The most common classical methods of measuring elemental abundances rely on the measurement of the equivalent width of absorption lines with well-known atomic parameters. Equivalent width measurements can then be placed on a curve-of-growth, given a stellar model, in order to derive the abundance of the species giving rise to the observed feature. One must also know the effective temperature and surface gravity, as these parameters have large and complicated effects on the strengths of lines. Frequently the effective temperature is determined by independent (and not necessarily self-consistent) means, e.g., by color-temperature relations \citep[e.g.,][]{cas11},  by imposing the excitation equilibrium or by first fitting the full spectrum with a subset of main stellar parameters and elemental abundances \citep[e.g.][]{gar16}.

Photospheric lines are broadened by various processes, including pressure broadening, rotation, and macro-/micro-turbulence. These sources of line broadening, combined, are typically of the order $v_{\rm broad} \simeq 1-10\,$km/s, which translates to an intrinsic spectral resolution of star of $R=\lambda/\Delta \lambda = c/v_{\rm broad} \simeq 10^4-10^5$. In order to resolve spectral lines one would therefore want to obtain spectra at $R \gtrsim 10^4$. At lower resolution the lines blend together, at least in cool and metal rich stars (the most common stars in most large stellar spectroscopy surveys). It can be difficult to measure elemental abundances through equivalent widths or line profiles of individual unblended lines. The most straightforward way to make progress in such cases, while preserving the full information content of the spectra, is to self-consistently fit entire spectral regions, which is the approach taken here.

Another advantage of operating at high-resolution is that one can isolate and focus on the spectral lines whose atomic parameters are well known from laboratory work, and discard spectral regions that are not well-modeled, as a way to mitigate systematic uncertainties in the models. This is a clear advantage of working at high-resolution \citep[but see also][]{cze15}, although the extent to which this issue can be mitigated or addressed at low-resolution has not been thoroughly addressed.

Finally, there are very subtle effects in the spectra of stars that would be entirely invisible at low-resolution, such as isotope ratios. For example the measurement of the $^{24}$Mg/$^{25}$Mg isotopic ratio, which induces a shift of $\sim 0.01\,$nm in the MgH spectral lines, or the $^6$Li/$^7$Li isotopic ratio \citep{lind13}, which requires 3D NLTE models to properly model the line shapes and derive reliable isotopic ratios.

%
%
%
%
%
%

\subsection{Quantifying information content with gradient spectra}
\label{sec:gradient-spectra}

In order to understand why low-resolution spectroscopy could possibly perform comparably well, we need a metric for the theoretically achievable uncertainties for each stellar label. A compact but mathematically rigorous way to do this is the Cramer-Rao bound \citep{cra45,rao45}, which we introduced in this particular context in \citet{tin16}. How well we can estimate a stellar label depends on two things, (a) how much a spectrum varies as we vary the stellar label, i.e., the response function of a spectrum, and (b) the flux uncertainties at each wavelength pixel and their covariances. Let $C$ be the covariance matrix of the normalized flux. The Cramer-Rao bound predicts that the covariance matrix of the stellar labels $K_{ij}$ can be calculated via

\begin{equation}
K_{ij}^{-1} = \overrightarrow{\nabla}_{\labels} f_{\rm model} (\lambda_1)_{i} \; C^{-1}_{\lambda_1,\lambda_2} \overrightarrow{\nabla}_{\labels} f_{\rm model} (\lambda_2)_{j}.
\label{eq:CR-calculation}
\end{equation}

\noindent
For each $(i,\lambda)$, the ``gradient spectrum'' $\overrightarrow{\nabla}_{\labels} f_{\rm model} (\lambda)_{i}$ measures the variation of the spectral flux at wavelength pixel $\lambda$ with label $i$. Eq.~\ref{eq:CR-calculation} then essentially takes the quadrature sum of the variations across different wavelength pixels, weighted by the uncertainties of the observed flux. If the spectral response to label changes is steep, we have large values for $\overrightarrow{\nabla}_{\labels} f_{\rm model} (\lambda)_{i}$ and hence small values for $K_{ij}$ -- more precise measurements. Similarly, if the observed spectra have a higher S/N, the values for $C$ will be smaller which will also result in smaller values for $K_{ij}$. The sum extends over the available pixels in the spectrum. Throughout this paper we assume that the wavelength sampling is always $\lambda/3R$ (we adopt a factor of 3, following the sampling of the APOGEE survey spectra).\footnote{Note that for most high-resolution echelle spectrographs the wavelength sampling and spectral resolution are not necessarily directly connected in this way.} We simplify the calculation in Eq.~\ref{eq:CR-calculation} by assuming that there are no correlations between adjacent wavelength points, i.e., $C^{-1}$ is a diagonal matrix. If the wavelength points are correlated, it is analogous to having fewer uncorrelated wavelength points. The absolute information content will decrease, but as we will show with different survey wavelength coverage (and hence different numbers of uncorrelated wavelength points), the relative information content between high- and low-resolutions will not alter much. Our conclusions which focus only the relative information content remain robust.

The covariance matrix $K_{ij}$ of the stellar labels and the gradient spectra are the quantities on which we base the majority of our results in this study. Not only it is a mathematically robust way to represent how much spectral information there is in the spectra, it also predicts which elements can be detected above a given significance threshold and the covariance between different stellar label estimates. Clearly, the calculation of $K_{ij}$ depends on the chosen resolution and wavelength range. As we vary the resolution, we will be summing up from different wavelength pixels, and the gradient spectra will also change with resolution. In short, in order to evaluate how low-resolution spectroscopy performs compared to high-resolution spectroscopy, we will study how $K_{ij}$ varies as a function of resolution, spectral type, and wavelength range.

%
%
%
%
%
%

\subsection{Many stellar labels from low-resolution spectra}
\label{sec:why-low-res}
We will now examine, at first qualitatively, how the uncertainties in stellar label estimates vary as a function of resolution in order to gain some basic intuition. It is qualitatively clear that at the same wavelength range and the same S/N per resolution element, a high-resolution spectrum must contain much more information than a low-resolution spectrum. But for spectroscopic surveys there are two important boundary conditions that need to be taken into account for a sensible comparison of spectra taken at different resolutions: the first one is the exposure time per object, which sets the survey speed; the second is the number of available detector pixels onto which each spectrum can be mapped. As detector ``real estate'' is an important boundary condition in highly multiplexing spectroscopic surveys, higher resolution generally forces the choice of a proportionally smaller wavelength range (assuming a fixed number of pixels per resolution element). As a consequence, the spectra at lower resolution will have higher S/N per resolution element and larger wavelength range (with a greater chance to enclose key diagnostic lines of different elements). Both effects work in favor of the low-resolution spectra. We can now evaluate analytically how $K_{ij}$ changes as we lower the resolution:

\begin{enumerate}
\item The {\it rms} depths of narrow spectral lines decrease inversely proportional to the width of the line spread function (LSF) kernel. As a result, the {\it rms} values of the gradient spectra scale as $R$. Seen another way, the equivalent width (the total integral) of spectral lines is constant at different resolutions, but the size of a resolution element is proportional to $1/R$. Therefore, the {\it rms} depth per wavelength pixel sampled at each resolution element must scale as $R$ so that the equivalent width (the sum of $\Delta$ resolution element width $\times$ gradient) is constant.

\item On the other hand, for fixed exposure time and object flux, the S/N per pixel will improve by $1/\sqrt{R}$ due to Poisson statistics.

\item Furthermore, for a given number of detector pixels, the wavelength range scales as $1/R$. Assuming the spectral lines are evenly distributed, we will collect $1/R$ times more spectral lines at low-resolution. As information adds in quadrature, having $R$ times more lines will improve the information content by a factor of $\sqrt{R}$, thus the precision improves proportional to $1/\sqrt{R}$.
\end{enumerate}

These simple arguments show that to first order low-resolution and high-resolution spectra should achieve the same uncertainty for stellar labels, given the sensible boundary condition of equal survey speed. In fact, provided that we have robust models and the ability to fit all stellar labels simultaneously, the uncertainty should be entirely independent of resolution, at least so long as the assumptions above hold. We will show with simulations in Section~\ref{sec:theoretical-study} that this insensitivity to resolution holds over a perhaps surprisingly large range in $R$. However in practice, the label uncertainty is not entirely independent of resolution, especially at the highest and lowest resolutions:

\begin{itemize}
\item For elements that have only few spectral lines, expanding the wavelength range does not necessarily generate more information. The newly included wavelength range might be devoid of spectral lines for some elements. So, for low-resolution spectra, we might lose information by a factor of $\sqrt{R}$. However, including a wider wavelength range also implies that low-resolution spectra can detect more elements that might have {\it no} detectable lines in high-resolution spectra of necessarily narrower wavelength range.

\item Once the LSF at very high-resolution becomes narrower than the intrinsic broadening of most lines, further increasing the resolution does not improve the gradient spectra. Therefore, for a fixed exposure time the information content of an observed spectrum will decrease at higher resolution. However, such high-resolution will in some cases be critically important for dealing with systematic issues, e.g., identifying and removing telluric features, which are intrinsically narrower than most stellar lines.

\item At very low-resolution (e.g., $R\sim 100$ as for Gaia's BP/RP spectra), estimates for stellar labels become more correlated. Mathematically, the covariance matrix $K_{ij}^{-1}$ becomes less diagonal.  In other words, once the estimates of different stellar labels become highly degenerate, their individual estimates become less precise.

\item When modeling low-resolution spectra one is forced to fit the full spectrum and one must therefore have knowledge of the line spread function (LSF) across a wide wavelength range. This can introduce additional challenges to measuring precise elemental abundances that are not as severe when modeling equivalent widths or line profiles of individual features from high-resolution data.
\end{itemize}

%
%
%
%
%
%

\subsection{Fitting multiple stellar labels simultaneously}
\label{sec:psm}

We have argued that low-resolution spectra contain the same amount of information for a fixed number of pixels and at fixed exposure time, but we can extract this information only if we are able to fit all stellar labels simultaneously. Generating state-of-the-art model spectra over a wide wavelength range takes several CPU hours for a given set of stellar labels. In a parameter space of 20--60 labels, it is computationally prohibitive to search for the best-fitting stellar labels through minimization -- each step in the minimization process will take several CPU hours. The standard approach to this problem is to create a synthetic library on an approximately rectilinear grid in the stellar label space, creating models at each grid point and then interpolating between them \citep[e.g.,][]{gar16}. However, in this method, the number of models needed grows exponentially with the number of labels, implying insurmountable computational cost for fitting 20--60 labels. We tackled this problem in \citet{rix16,tin16} and devised a new algorithm -- polynomial spectral models (PSM) -- that can fit 20--60 labels simultaneously. In essence, PSM constructs a model for the predicted flux at each wavelength point in the label space that is a polynomial function of all labels. But it does so in a way that requires only $N^2 \sim \,$1$,$000 {\it ab initio} models, for $N = \,$20--60 labels. The success of such modeling depends of course on whether the stellar spectra to be fit are well approximated by a PSM. In \citet{rix16}, we found that a single second order expansion captures almost all the label space spanned by the APOGEE sample of giants with $T_{\rm eff}> \,$4$,$000$\,$K. \citet{rix16} found that the median deviation of the normalized flux between {\it ab initio} calculated APOGEE models with $18$ parameters and the PSM models is only 0.001. Such an ``interpolation error'' is negligible as it is an order of magnitude smaller than the typical S/N of an observed spectrum (S/N$\,\simeq 100$). Furthermore, finding the best-fitting models with PSM is also extremely efficient because it regularizes the likelihood space in a $\chi^2$-minimization. For instance, we found that PSM fitting 100$,$000 APOGEE spectra with $20$ parameters requires less than $100$ CPU hours. PSM therefore appears to provide a practical solution to the requirement in low-resolution spectra of fitting all labels simultaneously. We will demonstrate how PSM can be used to fit low-resolution spectra in Section~\ref{sec:real-study}. 

\begin{figure*}
\vspace{-0.2cm}
\centering
$\begin{array}{c}
\includegraphics[width=1.0\textwidth]{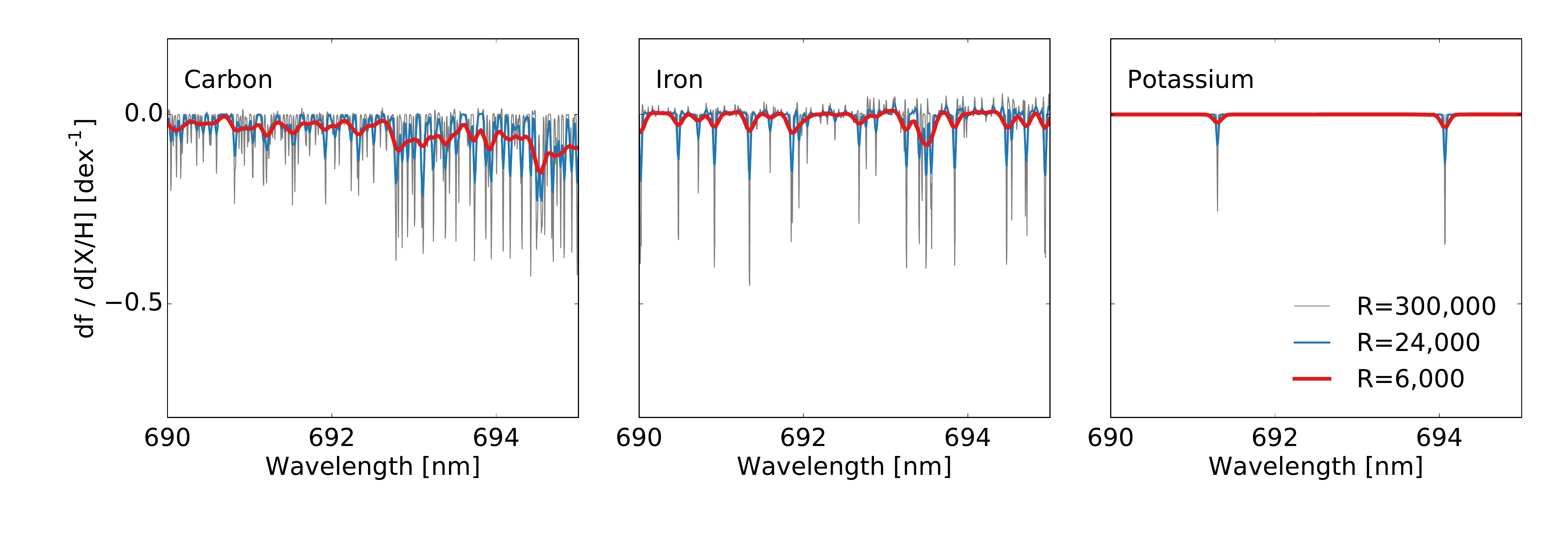}\\[-0.5cm]
\includegraphics[width=1.0\textwidth]{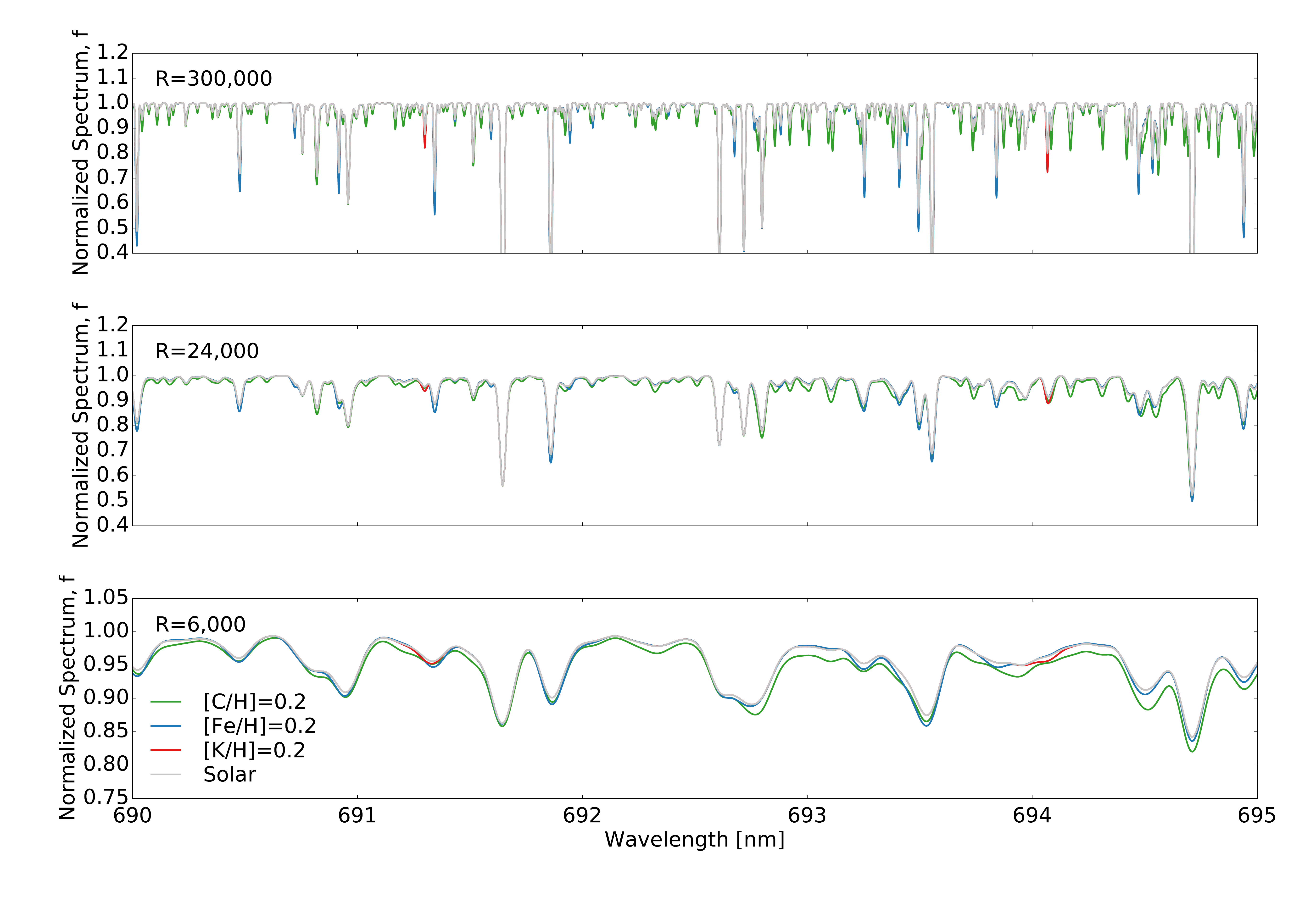}
\end{array}$
\caption{Illustration of the effects that changing abundances have on stellar spectra at different resolutions. The top panels show small wavelength segments of the gradient spectra of a solar metallicity, K-giant with respect to the abundances of C, Fe and K. We consider three different resolutions, $R = \,$300$,$000, 24$,$000 and 6$,$000. In the top panels, we calculate the gradient spectra by evaluating the difference between the solar spectrum and the spectrum with $\Delta [X/{\rm H}] = 0.2$ in each of these elements. The three lower panels show these enhanced normalized spectra (with respect to a solar metallicity spectrum). In each of these three lower panels, we show different spectra that are individually enhanced in C, Fe and K. At $R = \,$300$,$000 and $R =\,$24$,$000, some of the spectral lines remain unblended, however at $R=\,$6$,$000, all spectral lines are blended. Nonetheless, even at $R =\,$6$,$000, spectra that are enhanced in different elements show distinct features. But in order to extract elemental abundances at $R =\,$6$,$000, we must model the blended lines by fitting all stellar labels simultaneously.}
\label{fig:illustrate-gradient}
\end{figure*}

\begin{table*}
\begin{center}
\caption{Wavelength ranges, survey resolutions and approximation resolutions adopted in this study of the various surveys \label{table:coverage}}
\begin{tabular}{lcccc}
\tableline \tableline
\\[-0.2cm]
Survey & Wavelength range approximate (nm) & Survey resolution & Approximation resolution adopted here & Survey status
 \\[0.1cm]
\tableline
\\[0.0cm]
GALAH & 470--490, 565--585, 650--675, 760--790 & 28$,$000 & 24$,$000 & On-going\\[0.1cm]
APOGEE & 1$,$500--1$,$700 & 22$,$500 & 24$,$000 & On-going \\[0.1cm]
Gaia-ESO & 534--562, 848--900 & 20$,$000 & 24$,$000 & On-going \\[0.1cm]
4MOST (high-resolution) & 390--435, 515--575, 605--675 & 24$,$000 & 24$,$000 & Planned\\[0.1cm]
4MOST (low-resolution) & 390--885 & 6$,$000 & 6$,$000 & Planned \\[0.1cm]
WEAVE (low-resolution) & 370-1000 & 5$,$000 & 6$,$000 & Planned \\[0.1cm]
Gaia RVS & 840--880 & 11$,$500 & 8$,$000 & On-going \\[0.1cm]
RAVE & 840--880 & 7$,$000 & 8$,$000 & Completed \\[0.1cm]
SEGUE/BOSS & 390--900 & 2$,$000 & 2$,$000 & Completed \\[0.1cm]
LAMOST & 390--900 & 1$,$800 & 2$,$000 & On-going \\[0.2cm]
\tableline
\end{tabular}
\end{center}
\end{table*}

%
%
%
%
%
%

\section{The information content of low-resolution spectra}
\label{sec:theoretical-study}

In Section~\ref{sec:why-low-res} we discussed analytically, and qualitatively, why stellar parameter estimation should not depend strongly on spectral resolution under certain conditions. In this section we explore this issue in more detail by using synthetic model spectra and evaluating how uncertainties on stellar labels, calculated with Eq.~\ref{eq:CR-calculation}, vary as a function of spectral resolution. We use model spectra to calculate the gradient spectra, $\overrightarrow{\nabla}_{\labels} f_{\rm model} (\lambda)_{i}$, in Eq.~\ref{eq:CR-calculation} and the label covariance matrices that reflect the label uncertainties, under the assumption that the models are a good description of the data.

\subsection{Setup}

We compute 1D LTE model atmospheres from the {\sc atlas12} code maintained by R. Kurucz \citep{kur70,kur81,kur93}. We adopt the latest line list provided by R. Kurucz,\footnote{\href{http://kurucz.harvard.edu}{\tt http://kurucz.harvard.edu}} which include TiO, H$_2$O, CH, CN, CO, OH, MgH amongst other molecules. We evaluate the atmospheric structure with 80 zones of Rosseland optical depth, $\tau_{\rm R}$, with the maximum depth of $\tau_{\rm R}= \,$1$,$000. We automate the numerical convergence inspection for each calculated atmosphere and adopt the solar abundances from \citet{asp09}. We adopt the standard mixing length theory with a mixing length of 1.25 and no overshooting for convection. Spectra are evaluated with the radiative transfer code {\sc synthe} with a nominal resolution $R = \,$300$,$000 and are subsequently convolved to lower resolutions assuming a normal distribution with a FWHM of $\lambda/R$. 

To calculate approximate gradient spectra, we consider the differences of two spectra that differ by $\Delta T_{\rm eff} = 250\,$K, $\Delta \log g = 0.5$, $\Delta v_{\rm turb} = 0.5\,$km/s, $\Delta [X/{\rm H}] = 0.2$ with respect to a chosen reference point; \citet{tin16} elaborated why that is a sensible approximation. For any stellar label $\label_i$ and reference point, we calculate the gradient spectra as:
\begin{equation}
\nabla_{\labels} f_{\rm model} (\lambda,\labels_i) = \frac{f_{\rm model} (\lambda, \labels_i + \Delta \labels_i) - f_{\rm model} (\lambda, \labels_i)}{\Delta \labels_i},
\label{eq:gradient-calculation}
\end{equation}

\noindent
where $f_{\rm model}$ is the normalized flux of a model spectrum. In this study, we always perform full-consistent calculations -- we re-evaluate the atmospheric structure whenever we vary a stellar label, even though in many cases, e.g., for Eu, this is unnecessary \citep[see][for details]{tin16}. This is an important point because many elements have a significant effect on the atmospheric structure, which in turn can affect the emergent spectrum.  So for example an enhancement in Na not only affects the atomic Na {\sc i} lines but also, at a lower amplitude, large regions of the spectrum owing to the change in the atmospheric structure (Na is a major electron donor in cool stars).   

To study how the results vary for different stellar types, we consider a few reference points in this study, namely:
\begin{itemize}
\item M-giants: $T_{\rm eff} = \,$3$,$800$\,$K, $\log g = 0.5$
\item K-giants: $T_{\rm eff} = \,$4$,$800$\,$K, $\log g = 2.5$
\item G-dwarfs: $T_{\rm eff} = \,$5$,$800$\,$K, $\log g = 4.5$
\item F-dwarfs: $T_{\rm eff} = \,$6$,$800$\,$K, $\log g = 4.5$
\end{itemize}
\noindent
Note that since the reference points will be set at solar abundances or scaled-solar abundances, this study mainly investigates the spectral information content of a typical star with absolute abundances $A({\rm N}) < A({\rm C}) < A({\rm O})$, i.e., not a carbon star.

We adopt the following relation between $v_{\rm turb}$ and $\log g$ \citep{hol15}:\footnote{APOGEE calibrated this relation with giants, so this relation might not apply to the broad range of stellar types in this study. But the goal here is to have a wide variety of $\log g$ and $v_{\rm turb}$ as our reference points, so the exact relation between these two parameters does not impact our results.}

\begin{equation}
v_{\rm turb} = 2.478 - 0.325 \log g \,\,\,\,\, {\rm km/s},
\end{equation}

\noindent
The top panels of Fig.~\ref{fig:illustrate-gradient} illustrate gradient spectra for three  elements -- C, Fe, and K -- assuming a K-giant, solar metallicity reference point. We consider three different resolutions -- the nominal model resolution at $R = \,$300$,$000, a high-resolution mode, $R = \,$24$,$000, and a low-resolution mode, $R = \,$6$,$000. We also show the normalized spectra with and without enhancements in the abundances of these three elements in the lower panels. At $R = \,$300$,$000 and $R = \,$24$,$000, some of the spectral lines are resolved and unblended. These lines are typically selected to derive elemental abundances. Carbon has many more lines due to molecular contributions. Elements such as potassium have far fewer lines. However, at $R = \,$6$,$000, all lines are blended. If we wish to derive the elemental abundance of potassium, for example, we will need to model other elements contributing to the blends at the same time. Therefore, to extract spectral information at $R = \,$6$,$000, we need to model the blended lines by fitting all relevant stellar labels simultaneously. 

The top panels of Fig.~\ref{fig:illustrate-gradient} reveal a few interesting features. For e.g., at $R = \,$300$,$000, the global depths of spectral lines are not exactly 300$,$000$\,$/$\,$6$,$000$\,=50$ times deeper than $R = \,$6$,$000. There are three effects in play: (a) at $R = \,$300$,$000, the intrinsic broadening is larger than the LSF broadening. As we have discussed in Section~\ref{sec:why-low-res}, over-resolving lines does not improve the gradients. (b) When there are many overlapping lines, such as the carbon and iron lines, gradients do not degrade as much at low-resolution. One way to think of this is that overlapping/blended features have larger effective widths, so that convolution does not degrade the gradients in the same way as isolated lines. (c) Since we are convolving a spectral profile instead of a delta function, although the rms depth is proportional to $R$, the minimum point of the convolved profile alone does not necessarily scale exactly with $R$. The last effect has no influence on our arguments in Section~\ref{sec:why-low-res}, but the first two effects work in favor of low-resolution spectroscopy. They imply that the gradients only degrade linearly with $R$ at certain restricted conditions. For example, the potassium lines at $R=\,$24$,$000 and $R = \, $6$,$000 are less affected by these two effects and show a close-to-linear gradient degradation. But going from $R=\,$300$,$000 to $R=\,$24$,$000, especially for the carbon and iron lines, the gradients do not degrade proportionally with $R$.

Beside studying how the results vary for different stellar types, we also consider different wavelength ranges in this study. We will assume the wavelength ranges of the APOGEE, GALAH, Gaia-ESO\footnote{We assume the GIRAFFE HR10 and HR21 settings, with which most of the Gaia-ESO sample will be observed.}, 4MOST, WEAVE\footnote{Here we assume the low-resolution configuration of WEAVE. Note that WEAVE also plans to observe stars in a high-resolution mode but with a reduced wavelength coverage.} Gaia RVS, RAVE, SEGUE and LAMOST surveys. Their wavelength ranges and spectral resolutions are listed in Table~\ref{table:coverage} (and are visualized in Appendix~\ref{sec:information-contents}). Note that 4MOST plans to work at two configurations. The low-resolution configuration will operate on a larger wavelength range than the high-resolution configuration. We will show in the following subsections that, regardless of the wavelength range and stellar type, low-resolution spectroscopy can measure equally many elements with the same precision as high-resolution spectroscopy.

\begin{figure*}
\centering
\includegraphics[width=1.0\textwidth]{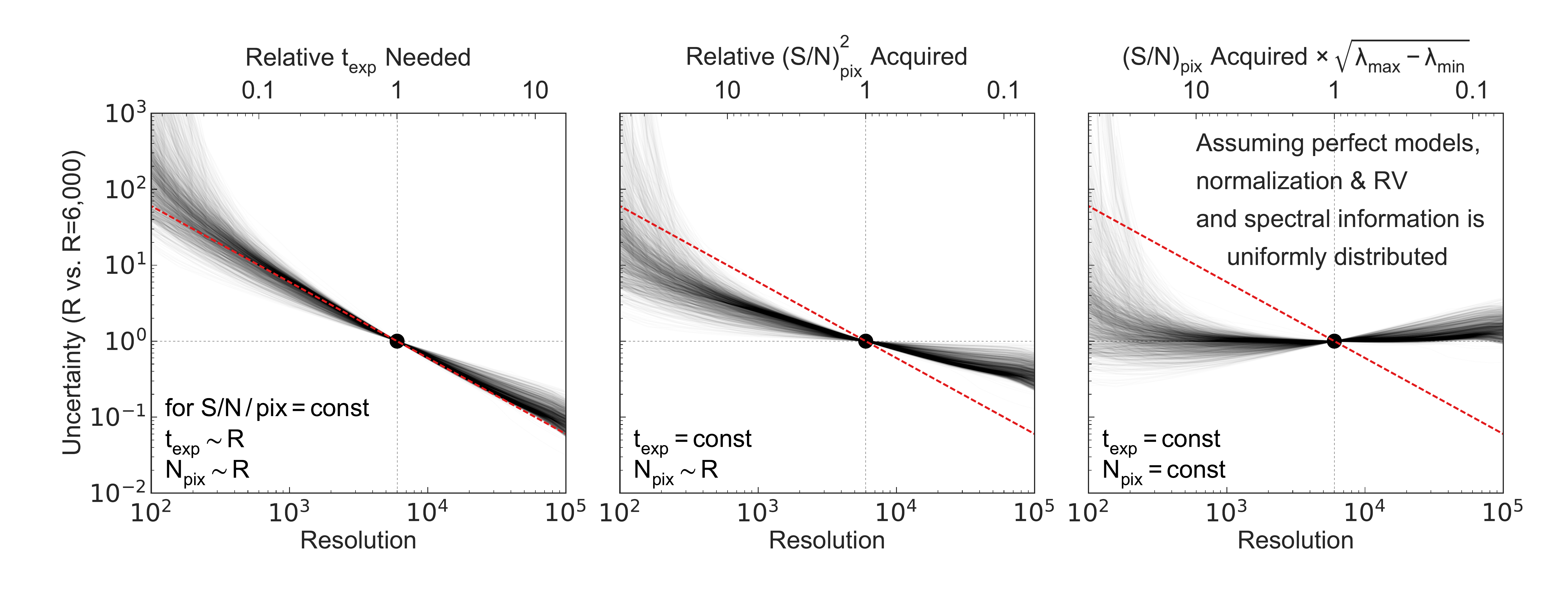}
\caption{Uncertainties of stellar labels as a function of spectral resolution, relative to $R= \,$6$,$000. We overplot results for all detectable stellar labels (stellar parameters and elemental abundances) from different stellar types, wavelength ranges and metallicities (see Section~\ref{sec:uncertainty} for details) because the result is general and independent of these choices. We assume perfect synthetic models, continuum normalization and radial velocity determination at all resolutions. We also assume that the information content is uniformly distributed throughout the entire wavelength coverage (see Section~\ref{sec:4most-survey} for caveats). The relative label uncertainties depend very much on the boundary conditions under which spectra of different resolutions are compared, as illustrated in the three panels. The left most one is a commonly used approach to such a comparison; the right most panel shows the comparison that is most pertinent to large spectroscopic surveys. Specifically, the left panel assumes that all resolution configurations have the same S/N per wavelength pixel (or resolution element). In this case, high-resolution spectra outperform low-resolution spectra, following a $1/R$ linear trend as depicted with the red dashed line, at the cost of significantly longer exposure times for higher resolution data. The middle panel assumes the same exposure time (and thus higher S/N for low-resolution spectra), and identical wavelength range (which would require $R$ times more detector real estate for high-resolution spectra). The right panel assumes the same exposure time and the same number of detector pixels (low-resolution spectra thus have wider wavelength range); as spectral diagnostic information is contained throughout the near-UV to near-IR spectra of the most common stellar types, broad wavelength range is very important. The right panel shows that, given the same exposure time and detector pixels, going to higher resolution (beyond $R \gtrsim \,$1$,$000) no longer improves the measurement uncertainties of stellar labels: the higher S/N per pixel, and the more extensive wavelength range compensate for lower spectral resolution, and vice versa. At very low-resolution with $R \lesssim $1$,$000, stellar label estimates become degenerate and deviate from the linear trend.}
\label{fig:no-gain-3}
\end{figure*}

%
%
%
%
%
%

\subsection{Stellar label estimates as a function of spectral resolution}
\label{sec:uncertainty}

In this section we evaluate how uncertainties of stellar labels vary as a function of spectral resolution, $R$, considering $T_{\rm eff}$, $\log g$, $v_{\rm turb}$ and all elements with atomic numbers from 3 to 99 as stellar labels, taking into account the correlations of the main stellar parameters such as $T_{\rm eff}$, $\log g$, [Fe/H], with other elemental abundances. We calculate the theoretical uncertainties of these stellar labels using Eq.~\ref{eq:CR-calculation}. The output covariance matrix $K_{ij}$ has the size of $100 \times 100$, showing the covariances of all stellar labels. The diagonal entries of $K_{ij}$ show variances ({\em marginalized over uncertainties of other stellar labels}) that one can achieve for each stellar label, and the square roots of these values give the theoretical uncertainties that we will explore in this section. Clearly, the gradient spectra depends on stellar type, wavelength range and metallicity. Therefore, we calculate $K_{ij}$ for different wavelength ranges, different stellar types and two metallicities -- $[Z/{\rm H}] = 0$ and $-2$. 

We also verified $K_{ij}$ by numerical simulations. We modify a reference spectrum with linear combinations of gradient spectra from all stellar labels and noise up the spectrum. We perform full spectral fitting (using PSM) via $\chi^2$-minimization and find that $K_{ij}$ gives the exact estimate of the covariance matrix of stellar labels. Finally, to study how theoretical uncertainties vary with $R$, we convolve gradient spectra to various resolutions, and recalculate $K_{ij}$ for each $R$. We define an element to be detectable if its uncertainty is better than $0.1\,$dex at $R = \,$24$,$000 and S/N$\,=100$.

Fig.~\ref{fig:no-gain-3} show the uncertainties as a function of $R$ of all detectable stellar labels (including $T_{\rm eff}$, $\log g$ and $v_{\rm turb}$). The figure overplots results from all stellar types, wavelength ranges and metallicities. We normalize the value of $y$-axis to be unity at $R= \,$6$,$000. Since uncertainty scales linearly with S/N (see Eq.~\ref{eq:CR-calculation}), the ratio of uncertainties plotted in Fig.~\ref{fig:no-gain-3} is independent of S/N. The left panel shows that, assuming the same S/N per pixel and the same wavelength range, the uncertainty degrades mostly linearly with $1/R$, regardless of stellar label, stellar type, wavelength range and metallicity, as explained in Section~\ref{sec:why-low-res}: since the absolute values of gradient spectra decrease proportionally to $1/R$, the uncertainties should also degrade linearly with $1/R$. 

However, given the same exposure time, low-resolution spectra will have a S/N per pixel that is  $1/\sqrt{R}$ higher than high-resolution spectra. In the middle panel, we take this into account and rescale the uncertainties in the left panel by $\sqrt{R/6,000}$. 

In the right panel, we also account for the larger wavelength range afforded by low-resolution spectra (given a fixed total number of pixels and a fixed number of pixels per resolution element) and further scale the uncertainties by another factor of $\sqrt{R/6,000}$, assuming spectral information distributes uniformly over the entire wavelength range. As expected from the arguments in Section~\ref{sec:why-low-res}, this factor compensates for the lower-resolution. {\em Remarkably, regardless of stellar type, wavelength range and metallicity, the achievable stellar label uncertainties are indeed nearly independent of spectral resolution, if we have robust models and can fit all stellar labels simultaneously (Fig.~\ref{fig:no-gain-3}, right panel).}

But Fig.~\ref{fig:no-gain-3} also quantifies how these simple trends are violated at both the very low-resolution and high-resolution ends. At the low-resolution end (e.g., $R \lesssim \,$1$,$000), stellar label estimates become nearly degenerate, resulting in reduced precision, as we have discussed in Section~\ref{sec:why-low-res}. As for the high-resolution end, spectral lines are eventually resolved, so further increasing the resolution does not improve the information content. By visual inspection, we found that for our model grid, most spectral lines  indeed have intrinsic broadening of the order of $R \simeq 10^{4.5}$. Over-resolving lines beyond this resolution does not improve the gradients and causes the high-resolution to deviate from the linear trend.

\begin{figure*}
\centering
\includegraphics[width=1.0\textwidth]{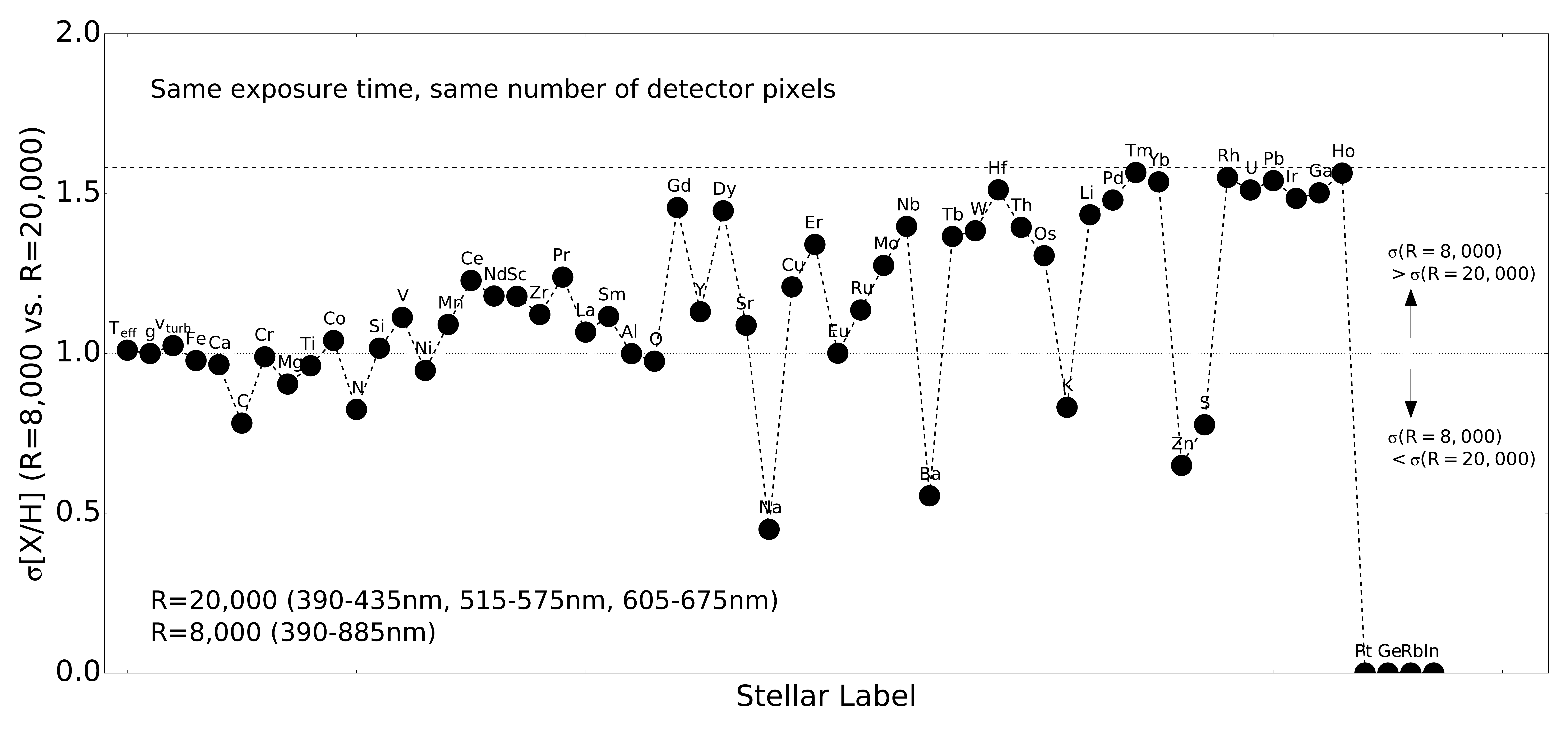}
\caption{Comparison of stellar label uncertainties for the two different resolution configurations of the 4MOST survey, assuming the same exposure time and number of detector pixels. We sort elements along $x$-axis according to their uncertainties for K-giants in the high-resolution configuration; we have also included three stellar parameters ($T_{\rm eff}, \log g, v_{\rm turb}$). For elements with numerous lines, such as Fe, Mg, the lower resolution appears fully compensated by the expanded wavelength range: high and low-resolution configurations perform equally well. For trace elements with only a few lines there are two regimes: if the signal comes from few, or even just one line, high-resolution spectra perform better, by up to the theoretical factor of $\sqrt{20,000/8,000}$ (the upper dashed line). In sharp contrast, there are elements where the wavelength range of the high-resolution configuration misses the (only) diagnostic lines (Pt, Ge, Rb, In); obviously, the low-resolution configuration performs far better in that case.}
\label{fig:no-gain-4}
\end{figure*}

\begin{figure*}
\centering
\includegraphics[width=1.0\textwidth]{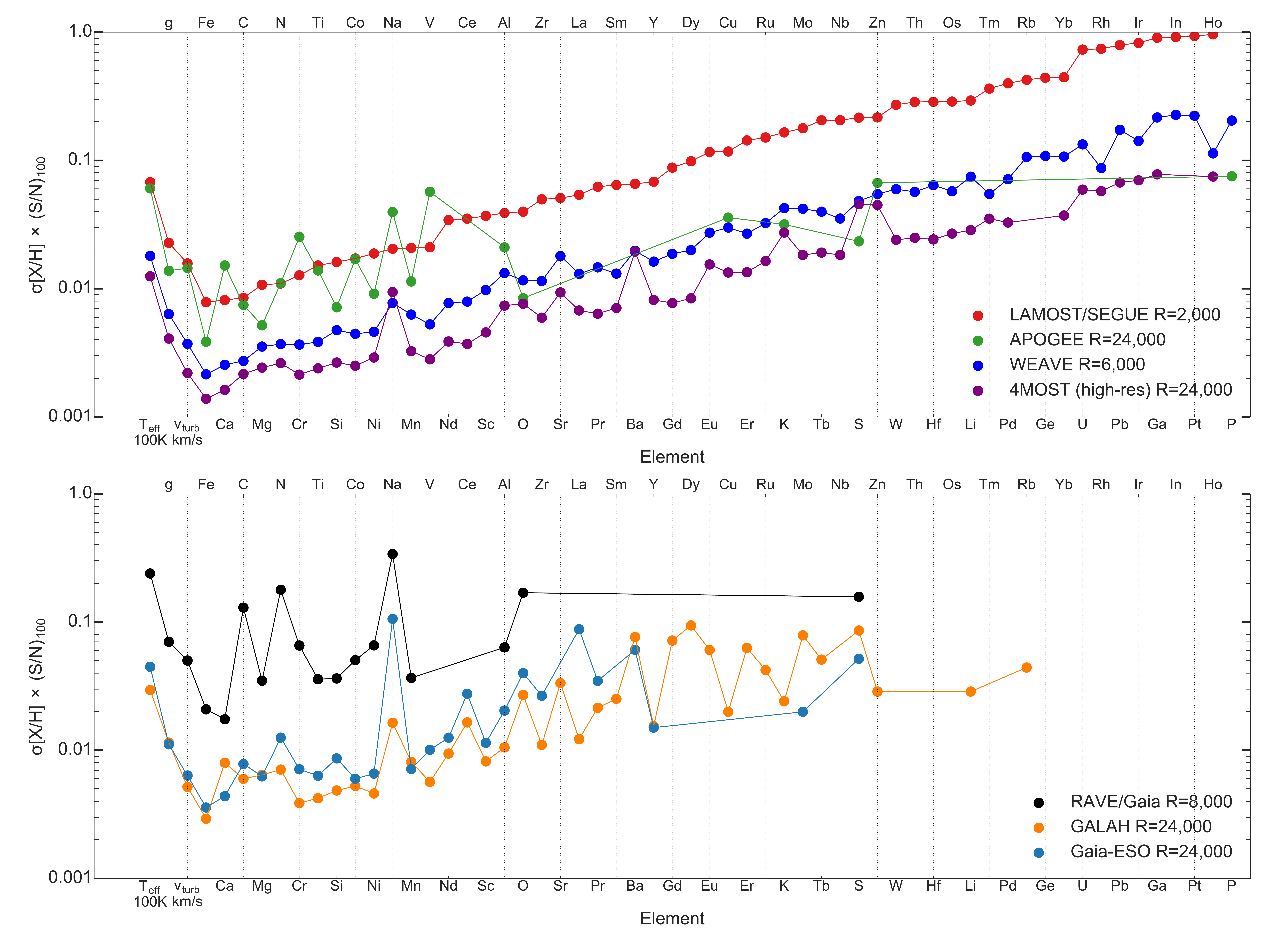}
\caption{Theoretical (best-case scenario) uncertainty of stellar labels for various spectroscopic survey configurations. Stellar abundances are sorted according to their uncertainties in low-resolution spectra ($R \simeq \,$2$,$000) that cover most of optical wavelength range (such as LAMOST, or SEGUE). For all surveys, we adopt the resolutions listed in Table~\ref{table:coverage}, solar metallicity, K-giants and a S/N per pixel $\, = 100$. We note that the precision might vary for different stellar types (see details in Section~\ref{sec:detectable-elements} and Appendix~\ref{sec:information-contents}.) If the synthetic models are robust, the $y$-values show the minimal uncertainties (Cramer-Rao bound) that we can achieve for stellar parameters and elemental abundances when fitting all stellar labels simultaneously. For elements where there are no useful spectral diagnostics in a survey's wavelength range the (filled) symbols have been omitted. Optical surveys like 4MOST, WEAVE, GALAH, Gaia-ESO, but also SEGUE and LAMOST can detect up to 50-55 elements and infrared surveys like APOGEE can detect up to 20 elements. More strikingly, if we have robust models, even with small wavelength ranges, RAVE and Gaia RVS can detect about 15 elements. The $(\mathrm{S/N})_{100}$ in the $y$-axis label is to remind that theoretical uncertainty scales linearly with S/N per pixel (cf. Eq.~\ref{eq:CR-calculation}), therefore for other S/N values, it suffices to scale the uncertainties as shown in the $y$-axis accordingly.}
\label{fig:elements-surveys}
\end{figure*}

\begin{figure*}
\centering
$\begin{array}{c}
\includegraphics[width=1.0\textwidth]{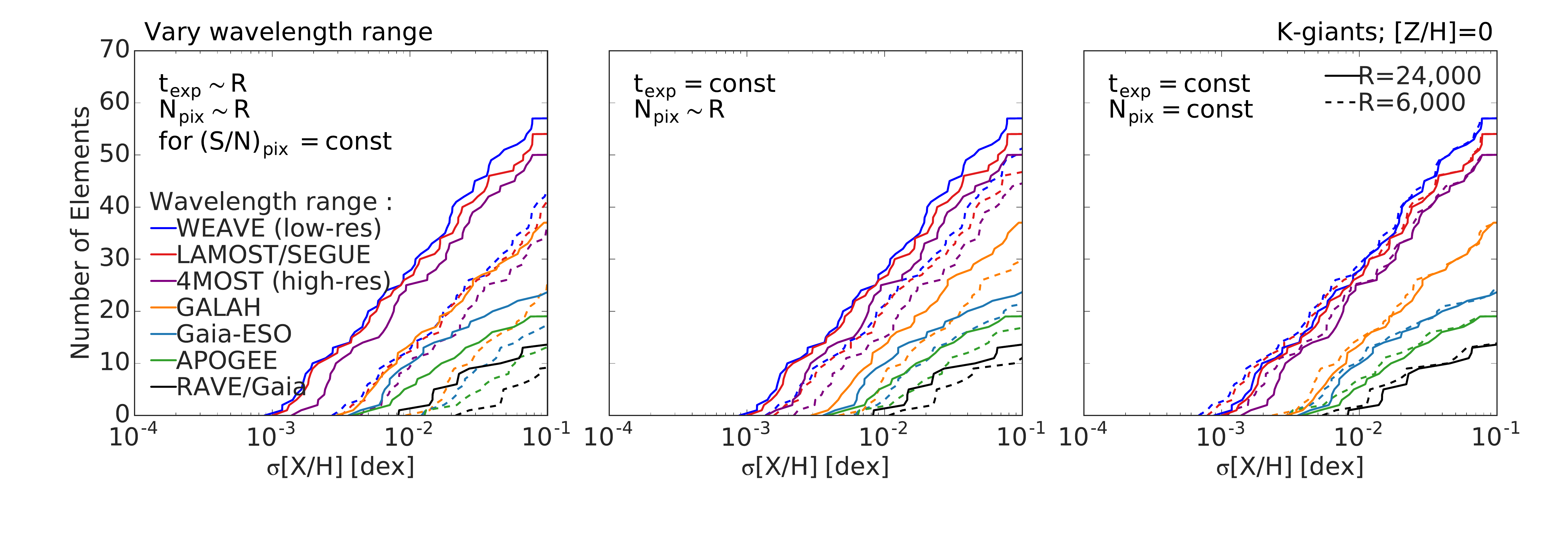}\\ 
\includegraphics[width=1.0\textwidth]{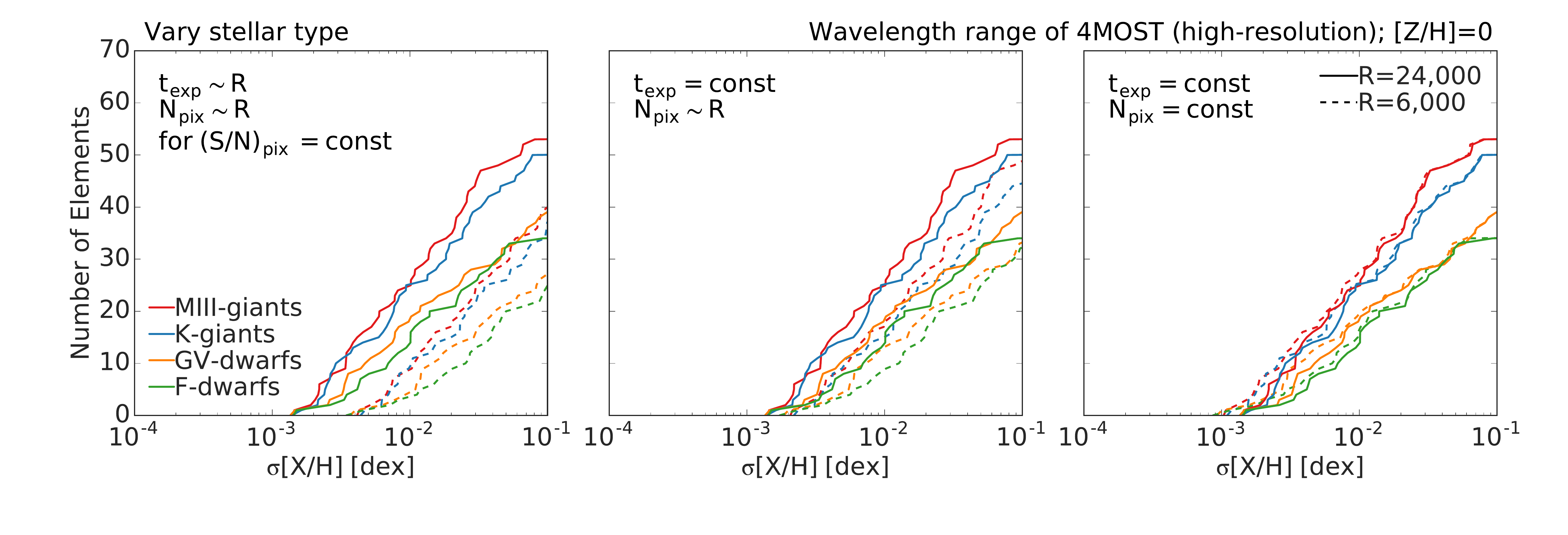}\\
\includegraphics[width=1.0\textwidth]{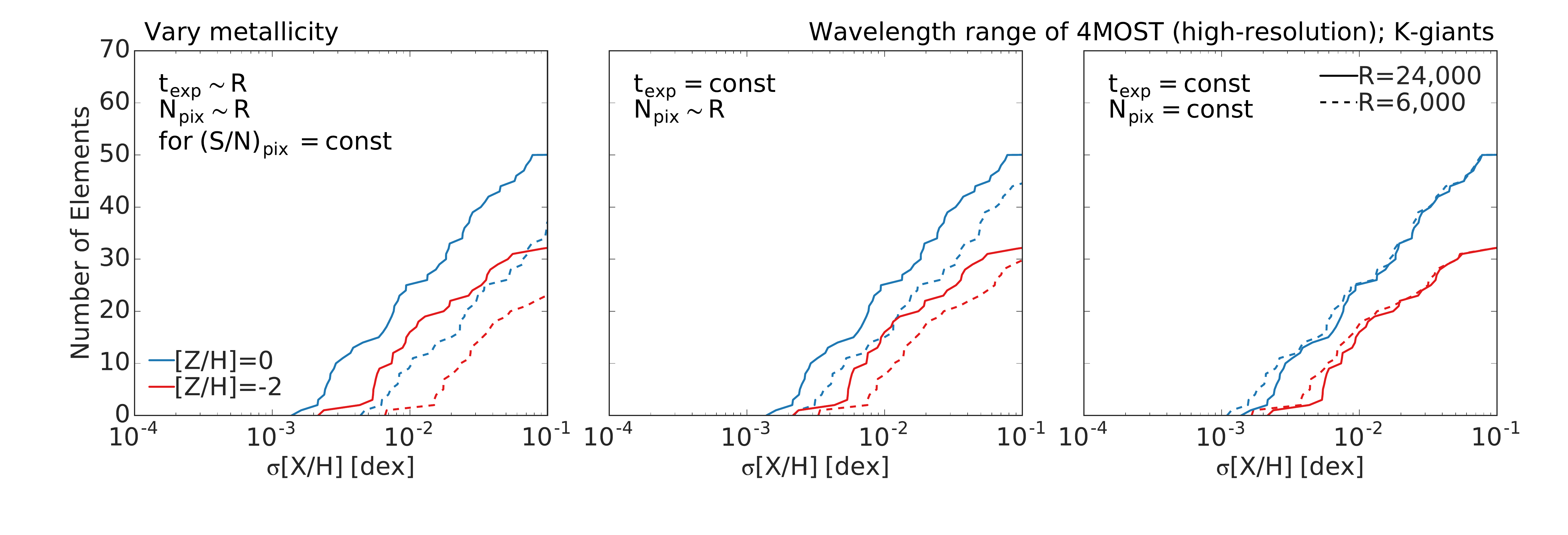} 
\end{array}$
\caption{The number of different elements ($y$-axis) for which abundances can be obtained with a certain (theoretical) uncertainty ($x$-axis), as a function of stellar type, wavelength range and metallicity. The solid lines assume a high-resolution configuration of $R = \,$24$,$000 and the dashed lines assume a low-resolution configuration of $R = \,$6$,$000. Panels from left to right illustrate three different comparisons between the high-resolution configuration and the low-resolution configuration. We assume a fixed S/N$\,=100$ per pixel and a fixed wavelength range for the high-resolution configuration, and vary the properties of the low-resolution configuration from left to right. The left panels assume the same S/N per wavelength pixel and the same wavelength range; the middle panels assume the same exposure time (higher S/N for the low-resolution configuration). The right panels further assume the same number of detector pixels (more extensive wavelength range for the low-resolution configuration), which we deem the most pertinent comparison. In that latter case, low-resolution spectra can detect as many elements as the high-resolution spectra with similar precision. Panels from top-to-bottom assume three different comparisons of survey targets and survey configurations: the top panels compare different wavelength ranges from various surveys; the middle panels compare different stellar types; the bottom panels compare stars with different metallicities.}
\label{fig:num-elements}
\end{figure*}

%
%
%
%
%
%

\subsubsection{4MOST survey as a case study}
\label{sec:4most-survey}

One can adopt simple arguments to rescale the uncertainties in Fig.~\ref{fig:no-gain-3} for a particular survey design. For example, one can assume that spectral line information is uniformly distributed throughout all wavelengths and derive an $\sqrt{R}$ improvement in uncertainty when going from the middle to the right panel of Fig.~\ref{fig:no-gain-3}. This assumption might be a good approximation for elements that have many spectral lines such as Fe, and $\alpha$-capture elements. But for trace elements, such as Li, K, that have only a few spectral lines, expanding wavelength range does not necessarily improve the information content. To work out a concrete example, we compare the two proposed resolution configurations of the 4MOST survey. Here we consider, for the same exposure time and a larger wavelength range, the tradeoffs in the low- {\it vs.} high-resolution setups for this particular survey.

4MOST proposes a high-resolution configuration with a shorter wavelength range of $390 - 435\,$nm, $515 - 575\,$nm, $605 - 675\,$nm and a low-resolution configuration with a wider wavelength range of $390 - 885\,$nm. These two configurations serve as a perfect case study to evaluate how uncertainty of stellar label changes when comparing low S/N high-resolution spectra spanning a narrow wavelength range, to high S/N low-resolution spectra spanning a large wavelength range. In Fig.~\ref{fig:no-gain-4}, we assume $R = \,$8$,$000 for the low-resolution configuration and $R = \,$20$,$000 for the high-resolution configuration. These resolutions are chosen such that both configurations consume an equal number of detector pixels when compensated with the difference in wavelength range. Also, for the same exposure time, the low-resolution configuration will have a higher S/N -- we assume the low-resolution configuration has a better S/N per pixel by a factor of $\sqrt{20,000/8,000}$.

Fig.~\ref{fig:no-gain-4} shows the ratio of uncertainties for all detectable elements and stellar parameters of the two configurations. Note that since we are plotting the ratio of uncertainties, the result is independent of the absolute values of S/N per pixel. In the $x$-axis, we sort elements by their uncertainties in the high-resolution configuration. If the two scaling relations as assumed in Fig~\ref{fig:no-gain-3} are exact, in particular, spectral line information is uniformly distributed throughout all wavelengths -- for e.g., information from stellar parameters: $T_{\rm eff}$, $\log g$ and $v_{\rm turb}$ -- the ratio should be close to unity. However if the information is concentrated only in a small wavelength range, expanding wavelength range does not collect more spectral information, and in this case, the low-resolution configuration will have a worse uncertainty by a factor of $\sqrt{20,000/8,000}$. The upper dashed line shows this value as the upper limit. Fig.~\ref{fig:no-gain-4} shows a clear trend -- stellar parameters and elemental abundances that have better uncertainties, such as Fe and Mg, generally have more lines, thus the uniform distribution of spectral information is a more valid approximation, resulting in ratios closer to unity. For elements that are less precisely measured, they are mostly elements that only have a small number of lines that reside in the wavelength region of the high-resolution configuration. Thus expanding the wavelength range at low-resolution does not help in this case.  Some elements (e.g., Na, Ba, K, Zn, S) are better measured (ratio$\, < 1$) at low-resolution. Expanding the wavelength range includes more lines from these elements that would otherwise be missed by the high-resolution configuration. The last four elements in Fig.~\ref{fig:no-gain-4} (Pt, Ge, Rb, In) highlight the scenario in which the high-resolution configuration does not cover any transitions of these elements, and so these elements are unmeasureable for this particular high-resolution configuration.

Although some elements perform worse at low-resolution even with the same exposure time and number of detector pixels, note that the ratio of uncertainties is bounded by an upper limit of $\sqrt{20,000/8,000}$. We can compensate this loss if we spend 20$,$000$\,$/$\,$8$,$000$\, = 2.5$ times more exposure time with the low-resolution configuration. Since low-resolution spectrographs are generally more accessible and high-resolution spectrographs have other downsides, such as lower instrumental throughput and more restrictive read noise limitations for faint targets, (see discussion in Section~\ref{sec:discussions}), it would still seem that low-resolution stellar spectroscopy with $R \simeq \,$6$,$000 and properly chosen wavelength range is the optimal strategy to design large-scale stellar spectroscopic surveys.

%
%
%
%
%
%

\vspace{1cm}
\subsection{Number of detectable elements for various surveys}
\label{sec:detectable-elements}

In the previous section we discussed how the ratio of uncertainties vary as a function of spectral resolution. In this section, we will study the absolute uncertainties -- i.e., the square root of diagonal entries of $K_{ij}$ in Eq.~\ref{eq:CR-calculation} -- given a fixed S/N per pixel, and determine how many elements we can, in principle, detect for various surveys. We assume a S/N per pixel of 100 in this section. We do not show the other values of S/N because the uncertainty scales linearly with S/N (cf. Eq.~\ref{eq:CR-calculation}).  We emphasize again that these uncertainties can only be attained if the model spectra are perfect, or nearly so. 

Fig.~\ref{fig:elements-surveys} shows the theoretically achievable uncertainties of detectable elements and stellar parameters for various surveys, assuming solar metallicity and K-giants. For each survey setup, we assume the adopted resolutions as stated in Table~\ref{table:coverage}. Optical surveys like 4MOST, WEAVE, GALAH, Gaia-ESO, SEGUE and LAMOST can measure up to $50-55$ elements. Strikingly, even for low-resolution spectra like SEGUE and LAMOST that has only $R \simeq \,$2$,$000, in principle, we can still measure as many elements as high-resolution spectra, provided that we can fit all stellar labels simultaneously and have robust stellar models. 

Infrared surveys, such as APOGEE, contain less information (also see Appendix~\ref{sec:information-contents}) and can ``only'' detect up to $20$ elements, consistent with the APOGEE pipeline \citep{hol15,sds16}. Not surprisingly, given the same resolution, surveys that have larger wavelength ranges such as 4MOST have smaller uncertainties than surveys that have more limited wavelength ranges like GALAH and Gaia-ESO. But interestingly, even for small wavelength ranges and low-resolution spectra from RAVE or Gaia, we can, in principle, detect about 15 elements, at least for K-giants. Measuring multi-elemental abundances with RAVE and Gaia RVS is an important application of PSM that we are currently exploring. 

In particular, for Gaia RVS, we found that most of the spectral information for C and N comes from the CN features. The spectral information for O comes from the CNO equilibrium -- when there is more oxygen, more carbon will be locked up in CO instead of CN. Therefore, the elemental abundance of O changes the CN features as well. As a result, one might measure C, N, O either directly or indirectly from the CN features. But we note that CNO are not completely degenerate because there are a few atomic lines from CNO in the RAVE and Gaia-RVS wavelength range that break this degeneracy. Most notably, we have a CI line at 872.713nm, a few OI lines around 844.636nm and an NI line at 868.028nm (but there are other weaker lines). We also note that the calculation of CR bound already took this partial degeneracy into account, and that is why the uncertainties for C, N, O for Gaia RVS are not excellent despite there is a lot of spectral information from the CN features. How well we can measure the CNO from the combination of the molecular features and the (blended) atomic lines is yet to be probed in practice, but it will be an important application of the Payne. Fig.~\ref{fig:elements-surveys} also suggests that high S/N spectra, such as stacked spectra from LAMOST and SEGUE, could detect many more elements than are currently being measured.

Instead of plotting uncertainties of individual elemental abundances, we can also compress this information and plot the cumulative distribution of uncertainties for all elemental abundances, as shown in Fig.~\ref{fig:num-elements}. The $y$-axis of Fig.~\ref{fig:num-elements} shows the cumulative number of elements that we can detect that have smaller theoretical uncertainties than the threshold shown in the $x$-axis. We consider two resolution configurations -- a high-resolution configuration of $R = \,$24$,$000 and a low-resolution configuration of $R = \,$6$,$000. Each row in Fig.~\ref{fig:num-elements} has three separate panels, showcasing three different possible comparisons between the low-resolution and high-resolution configurations, the same way as Fig.~\ref{fig:no-gain-3}. To recap, the panels on the left assume the same S/N per pixel and the same wavelength range. The middle panels assume the same exposure time and the right panels further assume the same number of detector pixels. For the middle panels and the right panels, we rescale the uncertainties of low-resolution spectra in the left panels by a factor of $\sqrt{6,000/24,000}$ and 6$,$000$\,$/$\,$24$,$000, respective, following Fig.~\ref{fig:no-gain-3}.

The top panels of Fig.~\ref{fig:num-elements} show the number of detectable elements at $R = \,$24$,$000 and $R = \,$6$,$000 for various wavelength ranges, assuming solar metallicity and K-giants. For surveys that have these resolutions, such as 4MOST, WEAVE, GALAH and Gaia-ESO, these panels are just compact representations of Fig.~\ref{fig:elements-surveys}. But we caution that for surveys that operate at a much lower resolution, such as LAMOST and SEGUE ($R \simeq \,$2$,$000), results in the top panels might not be directly applicable -- these panels only show the number of detectable elements if LAMOST and SEGUE were to operate in $R = \,$24$,$000 and $R = \,$6$,$000. Not surprisingly, at a given resolution and assuming the same S/N, the top panels show that a larger wavelength range, such as LAMOST and 4MOST, can detect more elements. These panels also show that, generally speaking, optical wavelength ranges contain more information and can measure more elements than the infrared. But more important, as shown in the right panel, if we assume the same exposure time and the same number of detector pixels, the dashed lines coincide with the solid lines, showing that $R =\,$6$,$000 spectra can detect as many elements as the $R = \,$24$,$000 spectra, echoing our earlier conclusions. This conclusion also holds true for the other comparisons that will we discuss next.

Thus far, we have only discussed how the detectability of elements vary as a function of wavelength range. But the detectability also depends on stellar type and metallicity. The middle panels of Fig.~\ref{fig:num-elements} show the number of detectable elements for different stellar types, assuming a wavelength range of the 4MOST (high-resolution) survey and solar metallicity. These panels show that cooler stars (e.g., M-giants) can detect more elements than hotter stars (e.g., F-dwarfs). This result is not surprising because cooler stars have more spectral lines, especially contributions from molecular lines. In fact. M-giants almost double the number of detectable elements compared to F-dwarfs. Since part of these cooler features come from molecular contributions and noting the composite nature of molecular features, this demonstrates the importance of full spectral fitting over many stellar labels simultaneously, without which we will not be able to extract information from molecular lines. 

Finally, the bottom panels show the number of detectable elements in two different metallicity regimes, $[Z/{\rm H}]=0$ and $[Z/{\rm H}]=-2$, assuming the wavelength range of 4MOST (high-resolution) and K-giants. We calculate the $K_{ij}$ matrix using gradient spectra with respect to reference points at these two different metallicities. Metal-poor stars have smaller gradient spectra which in turns predict a smaller number of detectable elements. Nonetheless, for optical surveys like 4MOST, although the number of elements is more restricted at the metal-poor regime, the bottom panels show that we can still detect up to 30 elements. Studies in Appendix~\ref{sec:information-contents} also indicate that there is still sufficient spectral information at low metallicity in the optical wavelength. But spectral information is more limited in the infrared, Although not shown, we found that we can only detect about $5$ elements at $[Z/{\rm H}]=-2$ with an APOGEE-like setup.

\begin{figure*}
\centering
\includegraphics[width=1.0\textwidth]{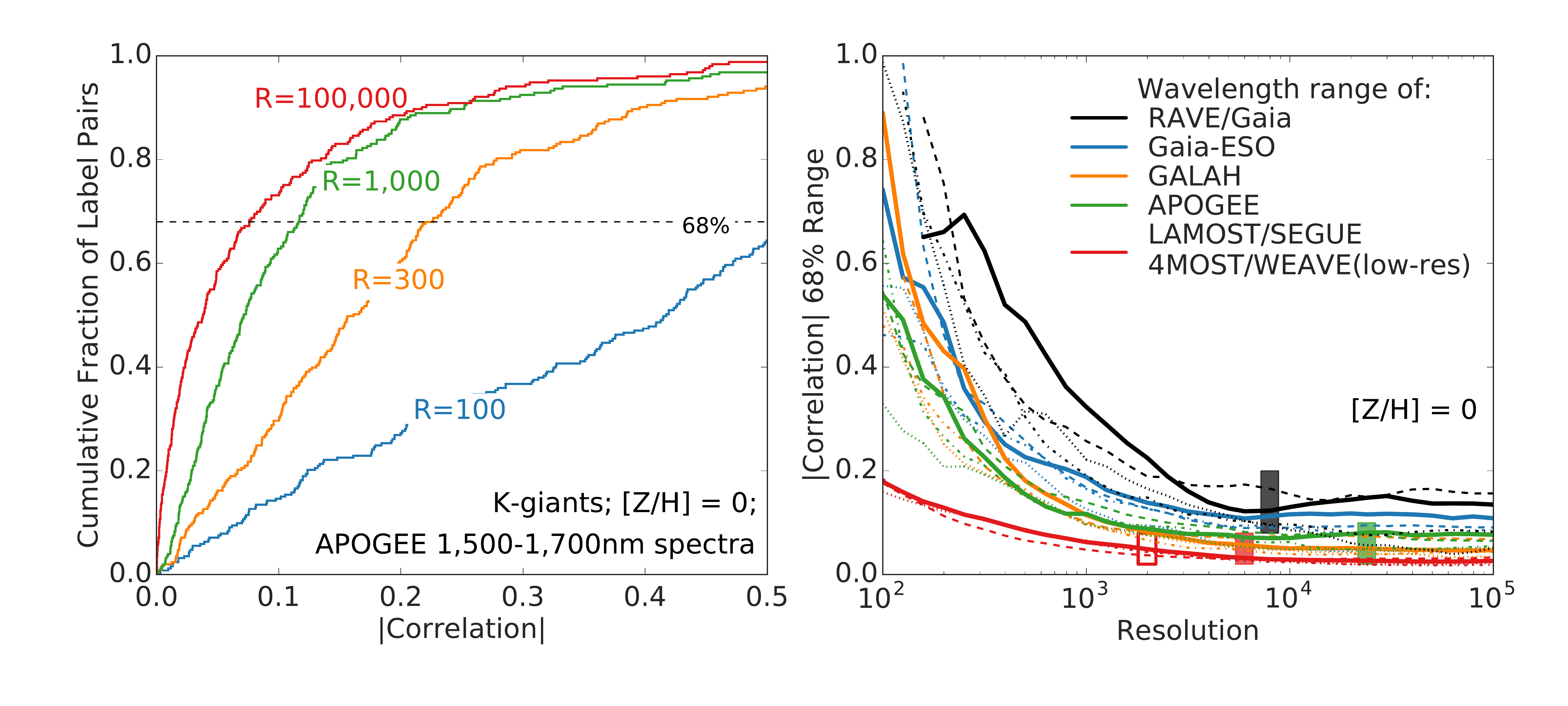}
\caption{Statistics of the correlation between the label estimates, as a function of spectral resolution and wavelength range. The left panel shows the cumulative distribution of correlations among all (detectable) label pairs. We assume a wavelength range of the APOGEE survey, solar metallicity and K-giants. The lines in different colors show the correlations assuming various spectral resolutions. Going from $R = 100$ to $R =\,$1$,$000 produces much more uncorrelated stellar label estimates yet, going from $R = \,$1$,$000 to $R = \,$100$,$000 barely reduces the correlations. The right panel shows how the typical level of label correlation ($68\%$ of pairs; see dashed line in the left panel) depends on the spectral resolution for the adopted wavelength range of different surveys (line colors). The right panel also illustrates the (weak) dependence of these correlations on spectral type: the solid, dashed, dashed-dotted and dotted lines assume stellar types of K-giants, M-giants, G-dwarfs, and F-dwarfs, respectively. Regardless of stellar type and wavelength range, the right panel shows that the label estimate correlations are generally modest, or even small, for $R \gtrsim \,$1$,$000; however, $\log g$ and $T_{\rm eff}$ are crucial labels that remain substantially correlated, even at high-resolution. The black box shows the survey resolutions of RAVE and Gaia; the green box indicates the resolutions of APOGEE, GALAH, Gaia-ESO and 4MOST (high-resolution); the shaded red box shows the resolution of 4MOST and WEAVE at low-resolution, and the hollow red box shows the survey resolutions of LAMOST and SEGUE. If one can fit all stellar labels simultaneously for these surveys, then most abundance estimates will not be seriously correlated, even though the spectral lines are blended.}
\label{fig:no-correlation-5}
\end{figure*}

\begin{figure*}
\centering
\includegraphics[width=1.0\textwidth]{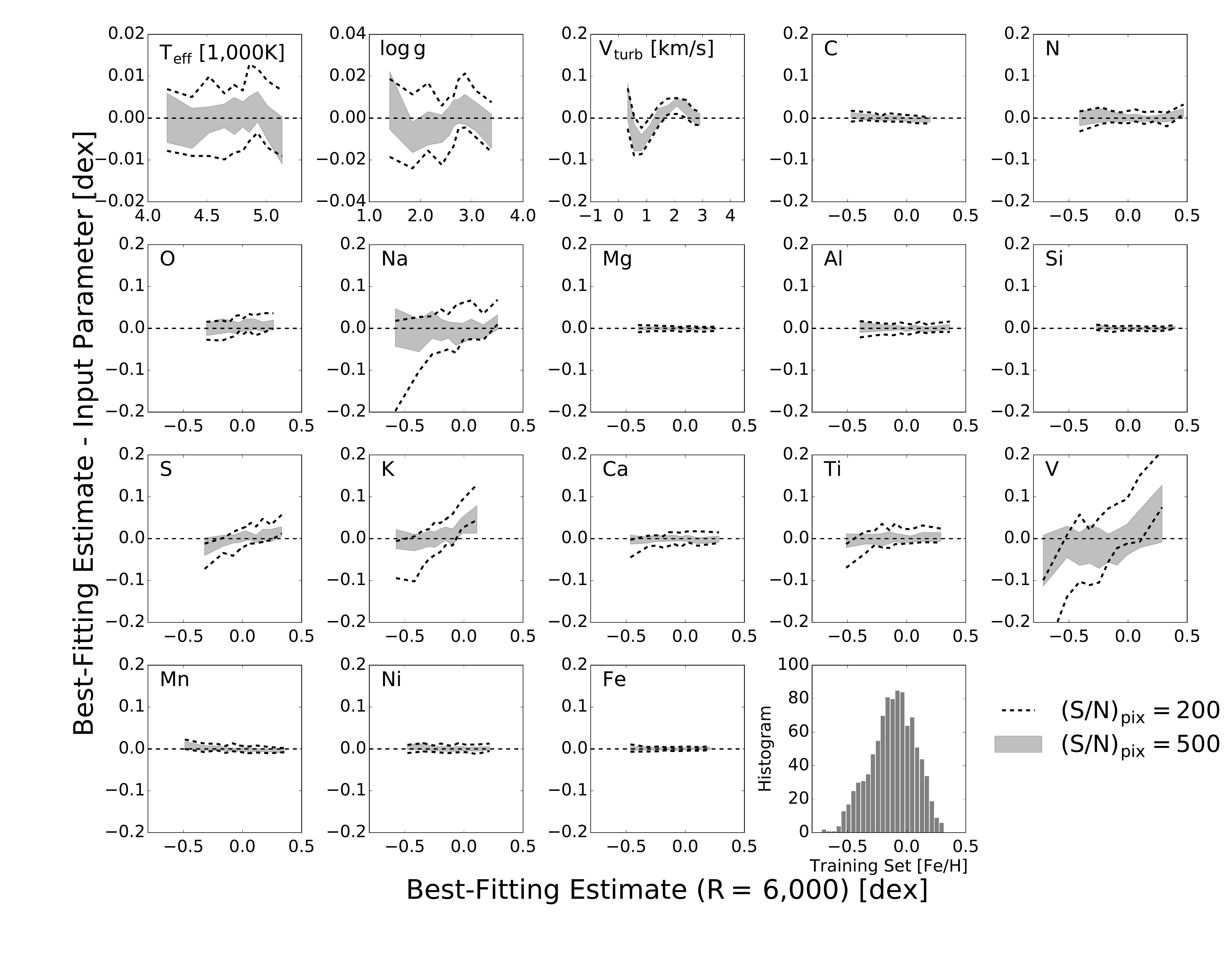}
\vspace{-0.5cm}
\caption{Stellar label precision resulting from the simultaneous fitting of all labels with a PSM model \citep{rix16} to spectra of APOGEE's wavelength range, but at a resolution of only $R=\,$6$,$000. We assume synthetic model spectra and adopt stellar labels from the APOGEE DR12 catalog. We consider stellar labels that have 4$,$000$\, {\rm K} \, < T_{\rm eff} < \,$5$,$500$\, {\rm K}$ and $1 < \log g < 4$, from which we generate 1$,$000 models to construct the PSM. The PSM is then used to fit another 1$,$000 testing models that have similar APOGEE stellar labels. Each panel shows the differences between the most likely PSM label estimates and the input labels. To mimic actual complications in spectral analyses, we also assume $10\%$ of the testing pixels to have large uncertainties, whose values are not used in spectral fitting. The gray band shows the $1\sigma$ range assuming S/N$\,= 500$ per wavelength pixel, and the dashed lines assume S/N$\,=200$ per wavelength pixel. Even with noised-up spectra and 10\% of bad pixels, almost all elemental abundances are recovered better than $0.01-0.05\,$dex from APOGEE-like spectra at $R = \,$6$,$000. However, we found that systematics in the PSM label estimates can become important at lower S/N; this is for elements that have a limited number of lines or only very weak signatures in the wavelength range, such as K, V or Na. For these elements, a non-parametric extension of the PSM model is needed to go beyond the simplistic quadratic assumption and improve the precision. The last panel shows the [Fe/H] distribution of the training spectra used in the construction of the PSM. Note that the $T_{\rm eff}$ and $v_{\rm turb}$ subplots assume different units than shown in the $y$-axis.}
\label{fig:apogee-low-res-1}
\end{figure*}

\begin{figure*}
\centering
\includegraphics[width=1.0\textwidth]{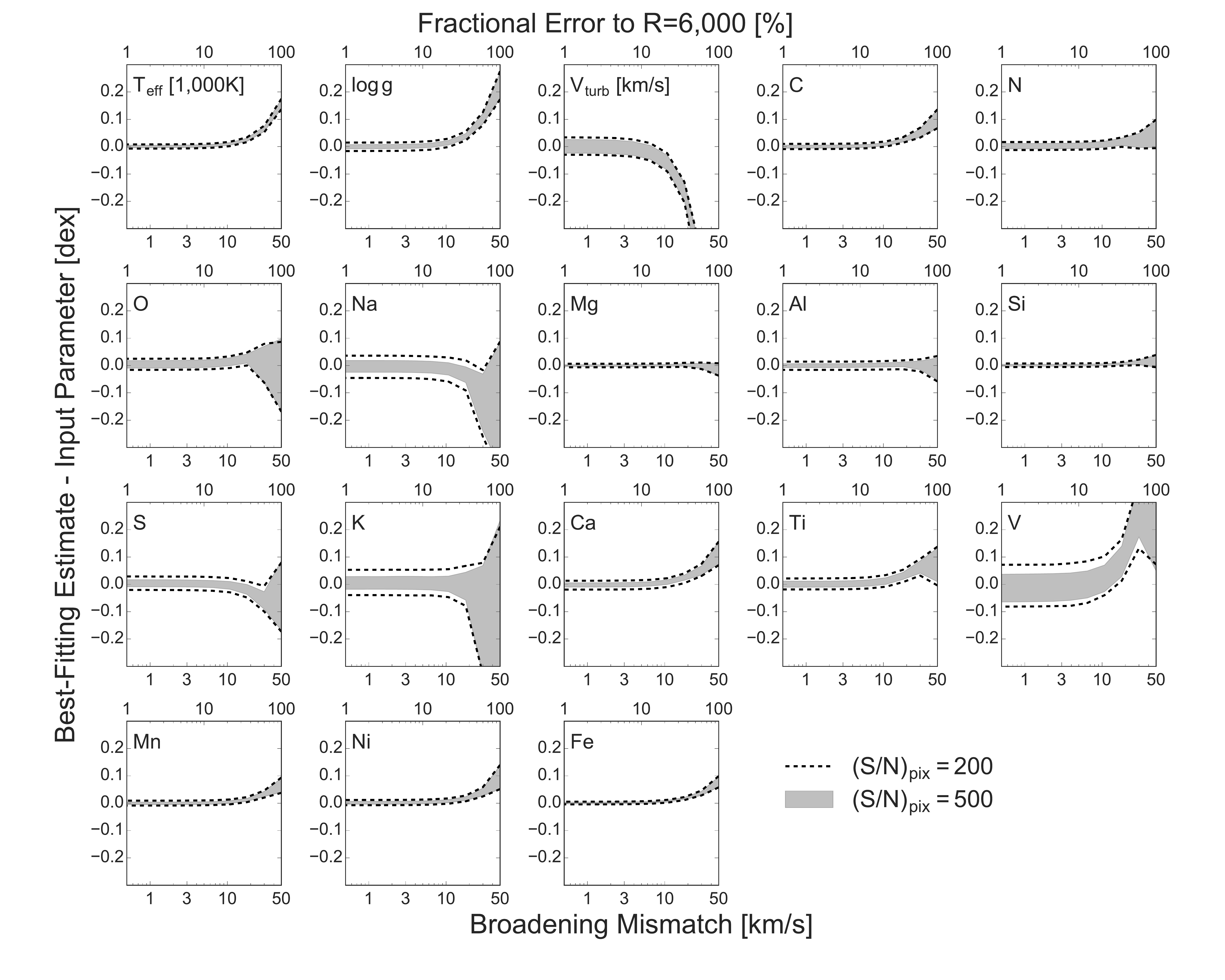}
\vspace{-0.2cm}
\caption{Sensitivity of low-resolution label estimates to mismatches in the assumed spectral line-spread function (LSF). The plot layout is the same as Fig.~\ref{fig:apogee-low-res-1}, but we study the deviation (averaging over the full range of the label) as a function of additional (and erroneous) LSF broadening in the PSM fitting. We also assume $10\%$ of bad pixels and S/N of 200 and 500 per wavelength pixel. The figure shows that the label estimates are insensitive to even substantive differences between adopted and true LSF. An additional of broadening of $< 10 \, {\rm km/s}$ has negligible effects on the estimates. In turn, this implies that low-resolution spectroscopy cannot recover $v_{\rm macro}$, modeled as a Gaussian convolution of the spectrum, for stars to the level of $\sim10$ km/s. But with a more severe additional broadening than $10 \, {\rm km/s}$, as the lines become shallower and broader we will overestimate $T_{\rm eff}$ and $\log g$, which in turn generally causes overestimations of $[X/{\rm H}]$ to compensate for the higher $T_{\rm eff}$.}
\label{fig:apogee-r-mismatch}
\end{figure*}

\begin{figure*}
\centering
\includegraphics[width=1.0\textwidth]{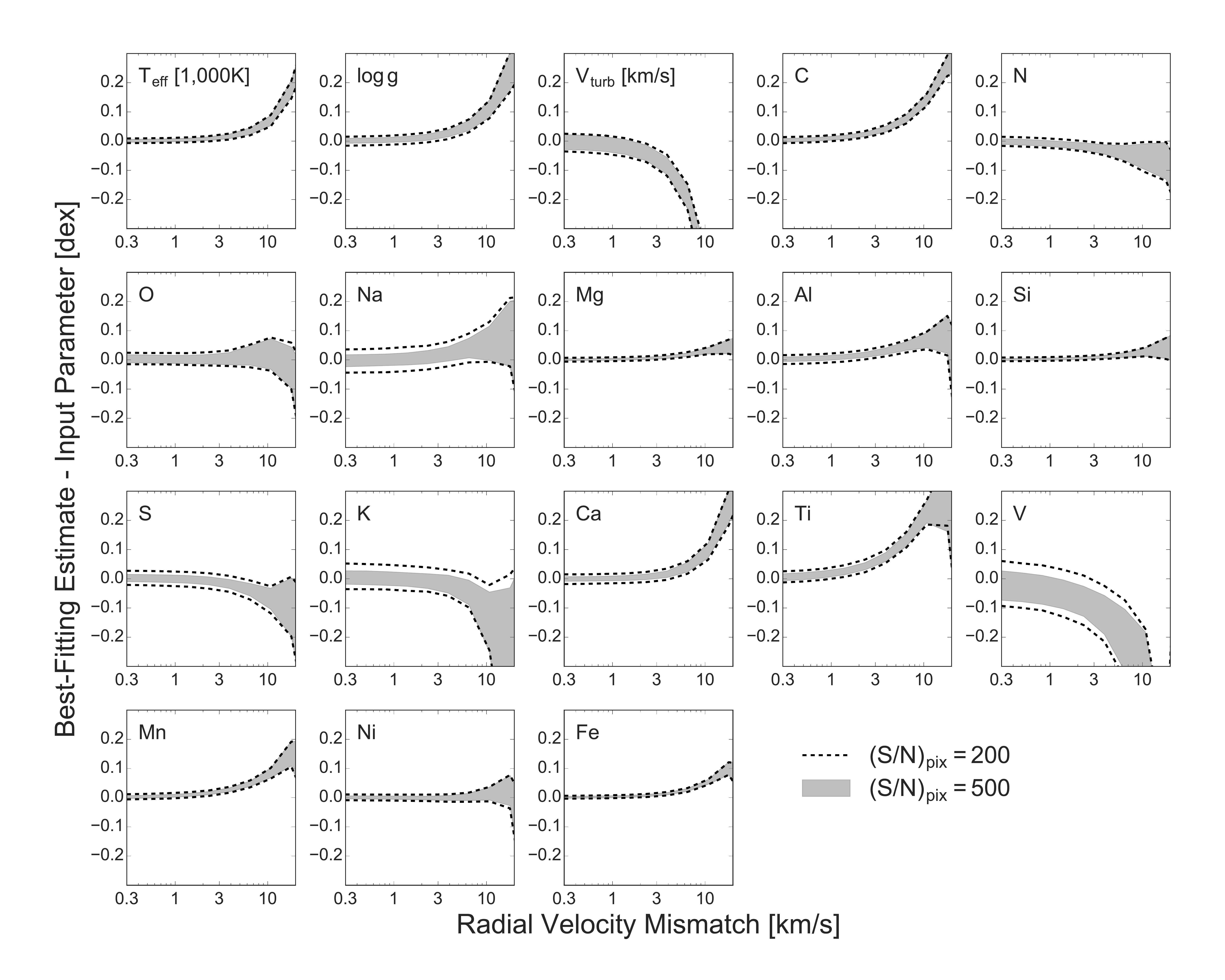}
\vspace{-0.2cm}
\caption{Sensitivity of low-resolution label estimates to mismatches in the assumed radial velocity. The plot is similar to Fig.~\ref{fig:apogee-r-mismatch}, but here we study the deviation as a function of additional (and erroneous) radial velocity. The figure shows that the label estimates are insensitive to an $< 10 \, {\rm km/s}$ error in radial velocity. The result demonstrates that although we might not be able to extract precise radial velocity from low-resolution spectra, the estimates of other stellar labels are largely unaffected by the less precise radial velocity measurements. However, this study in turn implies that low-resolution spectroscopy cannot recover radial velocity to the level of $\sim10$ km/s, which is one of the limitations of low-resolution spectra.}
\label{fig:apogee-rv-mismatch}
\end{figure*}

%
%
%
%
%
%

\subsection{Stellar parameter correlation as a function of spectroscopic resolution}
\label{sec:correlation-loss}

Thus far we have only considered the diagonal entries of $K_{ij}$ in Eq.~\ref{eq:CR-calculation} -- i.e., the theoretical uncertainties of stellar labels. However, there is more information in $K_{ij}$. This matrix also infers the correlations of stellar labels. More precisely, for each label pair $(i,j)$, the submatrix $\widetilde{K_{ij}}$ from the $(i,j)$ rows and columns of $K_{ij}$ shows the covariance of the $i$-th and $j$-th stellar labels, from which we calculate their correlation via 

\begin{equation}
C_{ij} \equiv \widetilde{K_{ij}}/\sqrt{\widetilde{K_{ii}}\widetilde{K_{jj}}}.
\end{equation}

\noindent
Uncorrelated estimations of stellar labels are crucial for Galactic studies because correlated estimates could make astrophysical interpretations difficult when looking for trends among stellar labels or searching for structures in chemical space \citep{tin15b}. In this section we will study how the correlation $C_{ij}$ varies as a function of spectral resolution.

The left panel of Fig.~\ref{fig:no-correlation-5} shows the cumulative correlations from all detectable stellar labels. The $y$-axis indicates the fraction of label pairs that have correlations smaller than the threshold value in the $x$-axis. We only consider the global distribution of correlations from all (detectable) label pairs in this section and refer interested readers to Appendix~\ref{sec:correlation-labels} for correlations of each label pair. The left panel shows that many label pairs have large correlations at $R = 100$ -- in fact, more than half of the label pairs have correlations larger than $0.4$. Strong degeneracies of stellar labels are expected at $R = 100$. At this resolution, there are only $30$ wavelength pixels in the APOGEE wavelength range, so most stellar labels are contributing to most of the pixels.  

However, increasing resolution to $R = \,$1$,$000 already removes or strongly diminishes the correlations between labels. About $80\%$ of the label pairs have correlations smaller than 0.15. In detail, as shown in Appendix~\ref{sec:correlation-labels}, only prominent stellar labels that contribute to most pixels, namely $T_{\rm eff}$, $\log g$, $v_{\rm turb}$, Fe, C, N, O are strongly correlated beyond $R =\,$1$,$000. The green and red lines in the left panel shows that going to a even higher resolution, such as $R = \,$100$,$000, no longer decreases the correlations by much. In practice, the correlations at low-resolution should be even smaller compared to high-resolution. Fig.~\ref{fig:no-correlation-5} assumes a fixed wavelength range, but as we have discussed earlier, given the same number of detector pixels, low-resolution spectra will have a much more extensive wavelength range, which allows further disentanglement of different contributions from various stellar labels.

The right panel in Fig.~\ref{fig:no-correlation-5} shows that this result is general and is independent of stellar type and wavelength range. Instead of plotting the cumulative distributions as in the left panel, the right panel plots the correlation values corresponding to the $68\%$ percentile of the cumulative distributions as a function of spectral resolution. The solid, dashed, dashed-dotted and dotted lines assume different stellar types -- K-giants, M-giants, G-dwarfs, and F-dwarfs, respectively, and the lines in different colors show results from various wavelength ranges. All these lines concur with the previous conclusion that stellar labels are not strongly correlated beyond $R \simeq \,$1$,$000, with the exceptions of the RAVE and Gaia RVS wavelength ranges. RAVE's and Gaia RVS's labels are only not strongly correlated beyond $R \sim \,$6$,$000 because RAVE and Gaia RVS have a limited wavelength range ($\lambda = 840-880 \,$nm). With this limited wavelength range, below $R \sim \,$6$,$000, there are too few wavelength pixels to distinguish contributions from various stellar labels.

The lines in the right panel show correlations at various resolutions, but for a spectroscopic survey, there is a well-defined survey resolution. An important question then is, are stellar labels correlated at the nominal survey resolutions? To answer this question, we label the survey resolutions with boxes in the right panel of Fig.~\ref{fig:no-correlation-5}. The black box shows the resolutions of RAVE and Gaia RVS. The green box shows the resolutions of APOGEE, GALAH, Gaia-ESO and 4MOST (high-resolution). The shaded red box shows the resolution of 4MOST/WEAVE in the low-resolution configuration; and the hollow red box shows the resolutions of LAMOST and SEGUE. All boxes are in the region where the correlation curves have already plateaued, indicating that stellar labels will not be strongly correlated from these surveys.

\begin{figure*}
\centering
\includegraphics[width=1.0\textwidth]{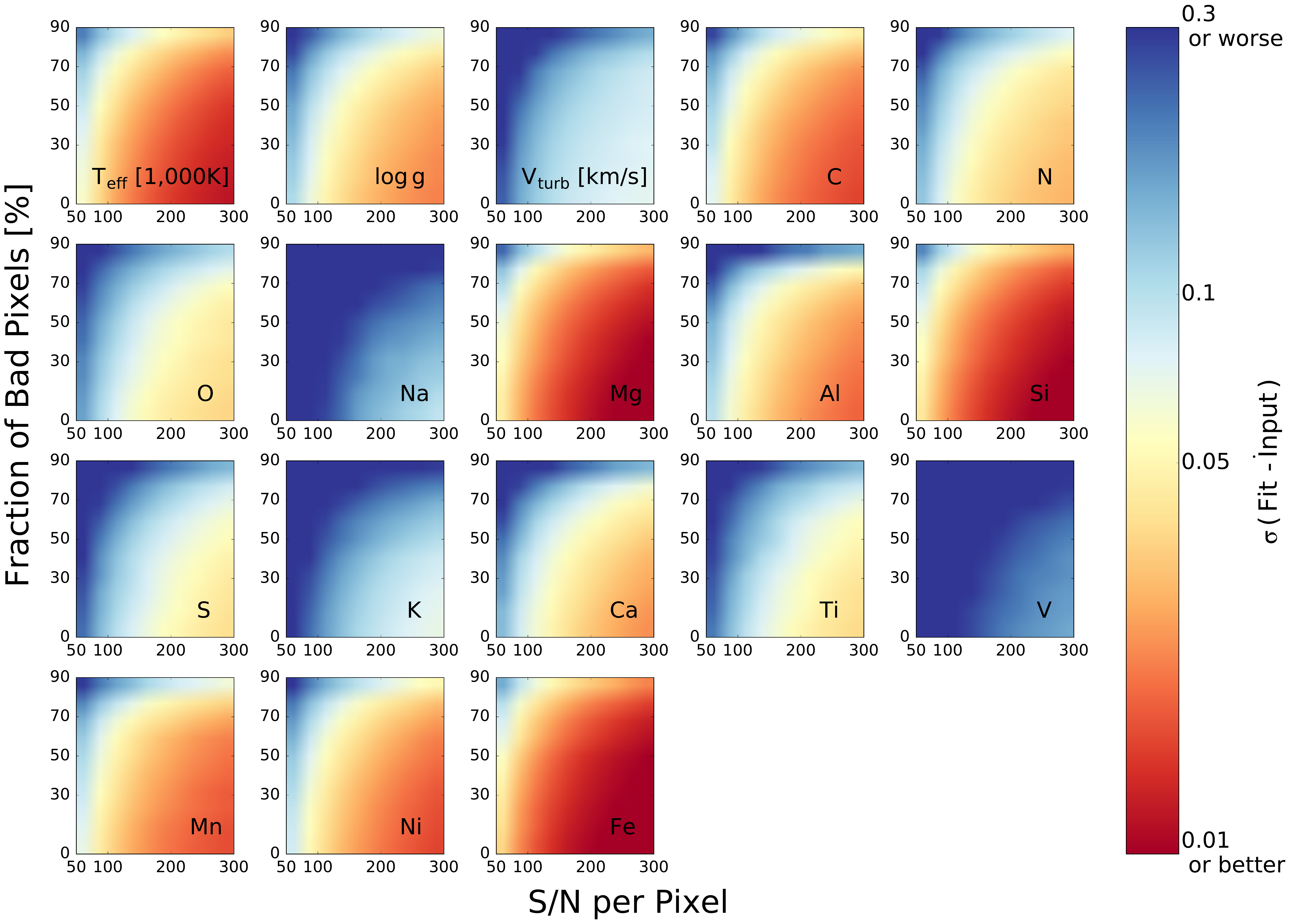}
\caption{Variance of different stellar label estimates obtained through PSM of APOGEE-like spectra, but presumed to be at $R= \,$6$,$000. The variances are calculated as a function of S/N per pixel and fraction of ``bad pixels''. At S/N per pixel $>100$, we can recover most stellar labels better than 0.1 dex. For stellar labels that have many spectral features across all wavelengths, such as Fe, Mg and Si, their recoveries depend less on the fraction of bad pixels. For stellar labels that have only weak gradients, such as $v_{\rm turb}$, V, Na, K, their recoveries are more compromised at $R = \,$6$,$000.}
\label{fig:apogee-low-res-2}
\end{figure*}

%
%
%
%
%
%

\section{Fitting mock spectra and deriving 18 stellar labels with $R= \,$6$,$000 spectra}
\label{sec:real-study}

Thus far we have shown that, given the same exposure time and the same number of detector pixels, spectral information remains largely independent of resolution. But there is still one crucial question yet to be answered, since most spectral lines at low-resolution are blended, can we model these blended features by fitting all stellar labels simultaneously? In other words, even though we know the spectral information is there, can we extract it? In this section we will generate and fit mock spectra at $R = \,$6$,$000 and show that we can recover multi-elemental abundances at this resolution, even in the presence of bad pixels, imperfections of LSF modeling, and flux uncertainties. 

We choose to study the wavelength range of APOGEE ($\lambda = \,$1$,$500--1$,$700$\,$nm) as our test case. We generate {\em flux-normalized} synthetic models at $R = \,$300$,$000 and subsequently convolve them to $R = \,$6$,$000 with a Gaussian kernel. We follow APOGEE DR12/DR13 and assume a wavelength sampling of $\lambda/3R$. With this sampling there are $\sim 1800$ wavelength pixels at $R = \,$6$,$000. Here we only study flux-normalized models and will discuss the potential problem with continuum normalization at low-resolution in Section~\ref{sec:limitations}.

We adopt the same PSM approach as in \citet{rix16} and perform full spectral fitting. Here we briefly summarize the idea of PSM. Instead of interpolating spectra, PSM constrains explicit quadratic functions that define how flux varies as a function of stellar labels at each wavelength. One can regard PSM as a second order Taylor expansion of a spectrum. How well PSM performs depends on the ``convergence radii'' of the Taylor sphere. The key to the success of PSM \citep[and related data-driven models such as the Cannon, cf.][]{nes15a,cas16} is that the Taylor sphere encompasses most of the stellar label space that matters -- i.e., the region of stellar label space where stars typically occupy. Previously in \citet{rix16}, we tested that we can fit all 18 stellar labels ($T_{\rm eff}$, $\log g$, $v_{\rm turb}$ and 15 elements) simultaneously and recover these stellar labels at $R =\,$24$,$000. Our aim here is to extend that analysis to $R = \,$6$,$000.

In order to test how well we can recover realistic stellar labels, we consider labels from the APOGEE DR12 catalog \citep{hol15}, restricting to objects with 4$,$000$\, {\rm K} \, < T_{\rm eff} < \,$5$,$500$ \, {\rm K}$ and $1 < \log g < 4$. We demonstrated that a single PSM region performs sufficiently well within this $T_{\rm eff}-\log g$ range for the high-resolution spectra in \citet{rix16}, and the results below will show that the PSM model works well for low-resolution spectra in this range as well. Going beyond this range might require multiple PSM spheres to cover the full relevant label space \citep[see][]{tin16}, but will not alter our conclusions. We also remove objects that have not measured abundances for all elements in the APOGEE DR12 catalog. We randomly choose 1$,$000 stars and use their stellar labels to generate our training set and constrain the PSM functions. We randomly choose another 1$,$000 stars to generate our testing set. The testing set is used to determined how well we can recover their input parameters. Note that to fully define a PSM for 18 stellar labels, we only need a minimal training set of $(18 \times 19)/2 = 171$ spectra. But \citet{rix16} found that overconstraining PSM with more training set, whenever it is still computational feasible, produces a better result. Therefore, we choose to constrain the PSM with 1$,$000 training models.

Fig.~\ref{fig:apogee-low-res-1} shows the recovery of input parameters for the testing spectra. Gaussian random errors are included assuming a S/N per pixel of $200$ and $500$. We also assign random values to $10\%$ of the testing spectra pixels and assign large uncertainties to these ``bad pixels''. These mimic pixels affected by skylines, cosmic rays or other possible contaminants. They also mimic pixels that are not well-modeled in real life applications and have to be subsequently clipped from spectral fitting. The gray shaded region in each panel illustrates the $1\sigma$ range and demonstrates that, with robust models, we can recover 18 stellar labels with a precision of $0.01-0.05\,$dex at $R = \,$6$,$000. The recovery of $v_{\rm turb}$ has a non-monotonic systematic. This suggests that the PSM model is not a perfect representation of the variation of flux as a function of $v_{\rm turb}$. A more complicated function might be needed to describe the variation, but even with a simple quadratic function, the systematic is small ($< 0.1\,$km/s). Similarly, we found that, for elements that have a limited number of lines or only very weak signatures such as K, V or Na, systematics in the PSM label estimates can become important at lower S/N. For these elements, a non-parametric extension of the PSM model is needed to go beyond the simplistic quadratic assumption and improve the precision. This is work that we are currently pursuing, but this technical limitation will not alter our conclusion that low-resolution spectra can be (nearly) equally information rich as their high-resolution counterparts.

Another potential source of systematic uncertainty is our imperfect knowledge of the line spread function (LSF).  To study how sensitive our results are with imperfect adopted broadening kernel, we model mock spectra that are further convolved with an additional broadening of $0.5-50\,$km/s. Fig.~\ref{fig:apogee-r-mismatch} shows the scatter between the best-fit and input stellar label as a function of additional broadening. As before, we also assume $10\%$ bad pixels and adopt S/N per pixel of 200 and 500. The figure shows that, at $R=\,$6$,$000, the estimates are not severely affected by an additional broadening $< 10\,$km/s. But if the LSF errors are larger than $10 \, {\rm km/s}$, the PSM estimates are biased. As spectral features become shallower and broader with additional broadening, we will overestimate $T_{\rm eff}$ and $\log g$, which in turn generally causes overestimations of $[X/{\rm H}]$ to compensate for the higher temperature. In contrast, although not shown, we checked that bad pixels (with large uncertainties assigned) and flux uncertainties, as studied in Fig.~\ref{fig:apogee-low-res-1} and Fig.~\ref{fig:apogee-low-res-2}, do not bias the stellar label estimates at high S/N. On the flip side, the weak dependence with additional broadening shows that we cannot measure $v_{\rm macro} < 10\,$km/s at $R =\,$6$,$000 because $v_{\rm macro}$ broadening is completely dominated by the instrumental LSF broadening at low-resolution. On the other hand, $\log g$ and $v_{\rm turb}$ can be recovered at low-resolution because their broadening effects are not simple convolutions with a kernel and so can be distinguished from the broadening due to the LSF.

Similarly, in Fig.~\ref{fig:apogee-rv-mismatch}, we study the deviation of stellar label estimates as a function of additional (and erroneous) radial velocity. For spectra at low-resolution, since most features are broadened and blended, it might be hard to measure radial velocity precisely through the Doppler shift of spectral lines. It is important to understand how much the stellar label estimates might be affected by erroneous radial velocities. Fig.~\ref{fig:apogee-rv-mismatch} shows that as long as our estimates of radial velocity are not off by $10 \, {\rm km/s}$ or more, the stellar label estimates are largely unaffected. On the other hand, since the deviation (and the $\chi^2$ of the spectral fitting) is not sensitive to a radial velocity shift of $< 10 \, {\rm km/s}$, the figure also implies that for low-resolution spectra at $R =\,$6$,$000, we will not be able to measure radial velocity to a precision better than $10 \, {\rm km/s}$, which is one of the limitations of low-resolution spectra. On the flip side, this study also illustrates that, at least for the S/N and wavelength adopted in Fig.~\ref{fig:apogee-rv-mismatch}, we could measure RV to the precision of about $10 \, {\rm km/s}$ with low-resolution spectra. But we note that the exact RV measurable from low-resolution spectra depends on the particular S/N and wavelength coverage in consideration.

Fig.~\ref{fig:apogee-low-res-2} shows the scatter in the stellar label recovery as a function of S/N and the fraction of assumed bad pixels.  With S/N per pixel $> 100$, PSM can recover most stellar labels better than 0.1$\,$dex, even with $\sim 50\%$ of bad pixels. The weak dependence with the fraction of bad pixels is not surprising -- since information only adds in quadrature, having a single reliable line can carry a lot of weight. Thus, for elements that have many spectral lines, such as Fe, Mg and Si, a high fraction of bad pixels does not substantially change the results. On the other hand, the measurement of stellar labels that only have weak gradients, such as $v_{\rm turb}$, V, Na, K, at low-resolution, can be seriously compromised by a large fraction of bad pixels and/or large flux uncertainties.

%
%
%
%
%
%

\section{Discussion}
\label{sec:discussions}

In this paper we have demonstrated that it is possible to derive precise many elemental abundances with low-resolution spectra if one is not limited by systematic shortcomings of spectral models. Perhaps more remarkably we show that -- at given exposure time and number of detector pixels -- low-resolution spectra can yield elemental abundances as precise as high-resolution spectra, and without strong correlations among stellar labels. In this section we discuss several important caveats to these conclusions and additional complications, both for high and low-resolution analyses.

%
%
%
%
%
%

\subsection{Some drawbacks to high-resolution spectroscopy}
\label{sec:throughput}

Practicalities aside, if there is little or no gain in label precision at a given survey speed between resolutions $R\sim \,$1$,$000 and $R\sim \,$100$,$000, one might then wonder what, if any, downsides exist to pursuing a survey at the upper end of this range? First, as discussed in previous sections, the general independence of elemental abundance precision on resolution assumes that information is spread uniformly throughout the spectrum. This is generally not the case, especially for important classes of elements such as r- and s-process neutron-capture elements. Because of this fact, there is some minimum wavelength range that is necessary to cover in order to probe a given set of elements. This fact would tend to work against collecting high-resolution data as multiple instrument configurations are required and hence the number of detector pixels required is not fixed but increases with resolution. 

Spectrographs with very high spectral resolution are not suited for even moderately faint objects: they tend to have lower throughput than low-resolution spectrographs, and the exposure time to overcome the read-noise dominated regime for faint objects is often prohibitive.

%
%
%
%
%
%

\subsection{Limitations of low-resolution stellar spectroscopy}
\label{sec:limitations}

The fundamental limitation in analyzing low-resolution spectra is the strong reliance on the {\it ab initio} models being of uniformly high quality over a large wavelength range. At high-resolution one can focus on lines with very accurate atomic data and that are known to form in relatively well-understood layers of the atmosphere. At low-resolution every ``feature'' is in reality a blend of many lines and so it is difficult to isolate the good from the bad regions of the model spectra. As a result, the fraction of wavelength pixels that are affected by strong lines increases at low-resolution, and strong lines usually suffer more from model imperfections.  The strong lines are especially sensitive to mircroturbulence, non-LTE, and treatment of the damping wings. Hence, the robust information content of low-resolution spectra can be adversely impacted due to model imperfections.

Deriving a robust model is clearly beyond the scope of this paper. Here we only aim to show that low-resolution spectra are remarkably information rich and contain information of many elemental abundances. However, it is worth emphasizing that a single (or several) bad pixels on their own will not necessarily compromise the fits at low-resolution. In most cases there is a vast amount of redundant information in the spectrum, and so even if for example one or more iron lines are in error, the many other good iron lines will dominate the determination of the final iron abundance. Also, data driven approaches such as {\it The Cannon} \citep{nes15a} can construct spectral models for low-resolution spectra (based on accurate and precise training labels, e.g., from high-resolution spectra) that are -- almost by construction -- without substantive systematic errors.

There are a variety of ways that one can mitigate the effects of model imperfections when fitting low-resolution spectra. At a minimum, one can (and should) fit ultra high-resolution spectral atlases of standard stars such as the Sun and Arcturus. The residuals in the fits to the standards can be convolved to low-resolution and used to down-weigh spectral regions that are poorly described by the models. A more ambitious approach would be to collect a sample of ultra high-resolution spectra and tune the models to fit those data (e.g., by astrophysically calibrating the atomic line parameters, which are often not known to high precision). Ideally the sample for which ultra high-resolution spectra are available should span the full range of parameter space that one is interested in studying at low-resolution. These tuned models should then by design provide excellent fits to low-resolution data.

There are other important aspects of fitting low-resolution spectra that are related to the data quality and characteristics. In principle, with perfectly flux calibrated data, one could choose to fit the fluxed spectrum directly. In practice spectra are often not flux calibrated to the precision required and so some methods for continuum normalization are adopted. At high-resolution one can either choose to measure equivalent widths or fit polynomials to regions of the spectrum that are free of (strong) absorption lines. At low-resolution there are no wavelength ranges that probe only the continuum, and so the method of normalization is more model-dependent. Nonetheless, \citet{cas16,ho16,ho17} have shown that even with relatively low-resolution spectra from RAVE ($R=8$,$000$) and LAMOST ($R=2$,$000$), the continuum normalization is an issue that can be overcome, at least with the data-driven method.

Accurate continuum normalization for the ab-initio fitting of low-resolution spectra is an important aspect that we are currently pursuing with observed spectra, but is beyond the scope of this paper. Here we propose three ways that might mitigate this problem and will them explore in future studies. (a) One can iterate the fits of the stellar labels (with PSM) and the continuum. This approach was adopted to deal with the low-resolution DEIMOS spectra and has proved to be robust enough to measure multiple elements with low-resolution spectra. (b) With the potential robust $T_{\rm eff}$, $\log g$ estimates from CMD fitting for billions of stars from Gaia, one might proceed by directly modeling the flux spectra (or at least the slope of the flux spectra) instead of the normalized spectra. (c) Since PSM allows us to fit for many parameters simultaneously, one can include the continuum polynomial coefficients as part of the PSM parameters, and fit the stellar labels and the continuum simultaneously.

Another advantage of high-resolution data is that it is much easier to subtract bright sky lines, which are intrinsically very narrow. This is more of a concern in the NIR where there is a forest of bright sky lines. Yet another advantage of working at high-resolution is that equivalent widths are independent of the LSF and so the precise wavelength-dependent instrumental resolution need not be modeled. At low-resolution the LSF must be accurately modeled in order to derive reliable parameters. In practice this means that a parameterized LSF should become part of the model.

Subtle effects such as asymmetric line profiles, due for example to 3D effects or spot modulation, and small shifts in line centers, due for example to isotopic ratio effects, Zeeman splitting, gravitational redshifting, or convective motions, are just several examples of effects that are unlikely to be detectable, even with perfect models, at low-resolution (detection of these effects with perfect models would also require perfect knowledge of the wavelength solution and LSF). Finally, although not directly related to deriving stellar parameters, another obvious advantage of high-resolution is precision radial velocity measurements (cf. Fig.~\ref{fig:apogee-rv-mismatch}), which have been instrumental to studying exoplanet populations.

%
%
%
%
%
%

\section{Conclusions}
\label{sec:conclusions}

Large spectroscopic surveys such as APOGEE, GALAH and Gaia-ESO are now collecting several orders of magnitude more stellar spectra in the Milky Way than all previous surveys combined. But the key to unravel the evolution of the Milky Way depends on how well we can turn stellar spectra into stellar labels -- stellar parameters such as $T_{\rm eff}$, $\log g$, $v_{\rm turb}$ and many elemental abundances. At resolutions below $R \lesssim \,$20$,$000 most spectral lines are blended. Deriving reliable stellar parameters therefore requires simultaneously fitting dozens of stellar labels in order to model the blended features. Fitting dozens of stellar labels simultaneously has only recently been demonstrated to be possible with the aid of polynomial spectral models (PSM). In light of this new technique, in this paper we explored how the  information content of stellar spectra varies as a function of resolution and explored the possibility of deriving multi-elemental abundances from low-resolution spectra. We emphasize that our conclusions only speak to the theoretical information content of low-resolution spectra. In reality, low-resolution spectra will be more limited by systematics of the spectral models as well as continuum normalization. Confirming our results with an analysis of observed low-resolution data will be an important next step. Our findings in this paper are summarized below:

\begin{itemize}

\item We explore the information content in spectra covering 300--2$,$400$\,$nm, considering different wavelength ranges of past, on-going, and future spectroscopic surveys -- APOGEE, GALAH, Gaia-ESO, 4MOST, WEAVE, Gaia RVS, RAVE, SEGUE and LAMOST -- and different stellar types, from M-giants to F-dwarfs. Assuming that the underlying models (whether {\it ab initio} or data-driven) are without systematic errors, we find that optical surveys can measure 50--55 elements, and infrared survey can measure about 20 elements, even with low-resolution, $R \simeq \,$6$,$000, high S/N spectra. Even smaller wavelength ranges associated with the RAVE and Gaia RVS surveys can potentially measure up to 15 elements at high S/N.

\item Assuming the same exposure time per star and same number of detector pixels (e.g., $R \cdot (\lambda_{\rm max} - \lambda_{\rm min}) = \,$constant, and a constant number of pixels per resolution element), the derived uncertainties on stellar labels are essentially independent of resolution for 1$,$000$\, \lesssim R \lesssim \,$100$,$000. 

\item Even though spectral lines are blended at low-resolution, most stellar labels are not correlated at $R \gtrsim \,$1$,$000. This holds generically for elements that produce detectable features at more than one location in the observed spectrum.

\item We demonstrate that it is possible to recover 18 labels from low-resolution $R = \,$6$,$000 APOGEE-like model spectra, even in the presence of a significant fraction of bad pixels, imperfections in modeling the LSF, and realistic observational uncertainties.

\item Deriving precise many elemental abundances from low-resolution spectra could open up new windows for Galactic archeology, and in particular, chemical tagging because the latter requires a vast sample size, which is generally more challenging to obtain at high-resolution. We suggest that, in order to optimize scientific returns, a strategy for future spectroscopic surveys would be to collect a small number of ultra high-resolution ($R \simeq \,$100$,$000) spectra for model calibration purposes but to carry out the main survey at much lower resolution.

\end{itemize}

%
%
%
%
%
%

\acknowledgments

We thank Bob Kurucz for developing and maintaining programs and databases without which this work would not be possible. We also want to thank Benjamin Johnson, David W. Hogg, Jon Holtzman, Melissa Ness, Rob Simcoe, Matthias Steinmetz, Pieter van Dokkum, David Weinberg, Martin Asplund and Daniel Weisz for illuminating discussions. Y.S.T. was supported by NASA Headquarters under the NASA Earth and Space Science Fellowship Program - Grant NNX15AR83H. C.C. acknowledges support from NASA grant NNX13AI46G, NSF grant AST-1313280, and the Packard Foundation. H.W.R.'s research contribution is supported by the European Research Council under the European Union's Seventh Framework Programme (FP 7) ERC Grant Agreement n.$\,$[321035]. The computations in this paper were partially run on the Odyssey cluster supported by the FAS Division of Science, Research Computing Group at Harvard University. The computational work also used the Extreme Science and Engineering Discovery Environment (XSEDE), which is supported by National Science Foundation grant number ACI-1053575.

\appendix

%
%
%
%
%
%

\section{Information content of stellar spectra}
\label{sec:information-contents}

In this section, we explore the total spectral information content as a function of wavelength by adopting the idea of gradient spectra, i.e., how much a spectrum changes as we vary elemental abundances. We calculate gradient spectra for elements with atomic numbers from 3 to 99 (Li to Es), from $\lambda = \,$300--2$,$400$\,$nm, at $R = \,$300$,$000, and $\Delta [X/{\rm H}] = 0.2$. For the purpose of illustration, the gradient spectra are subsequently boxcar-smoothed with a bin size of $10\,$nm. Despite exploring an exhaustive list of elements, we find many elements to have zero gradient spectra because there are no significant atomic lines for these elements. We compare the information content at two different metallicities -- $[Z/{\rm H}] = 0$ and $[Z/{\rm H}] = -2$, and four different stellar types -- M-giants, K-giants, G-dwarfs, F-dwarfs. 

\begin{figure*}
\centering
\includegraphics[width=0.75\textwidth]{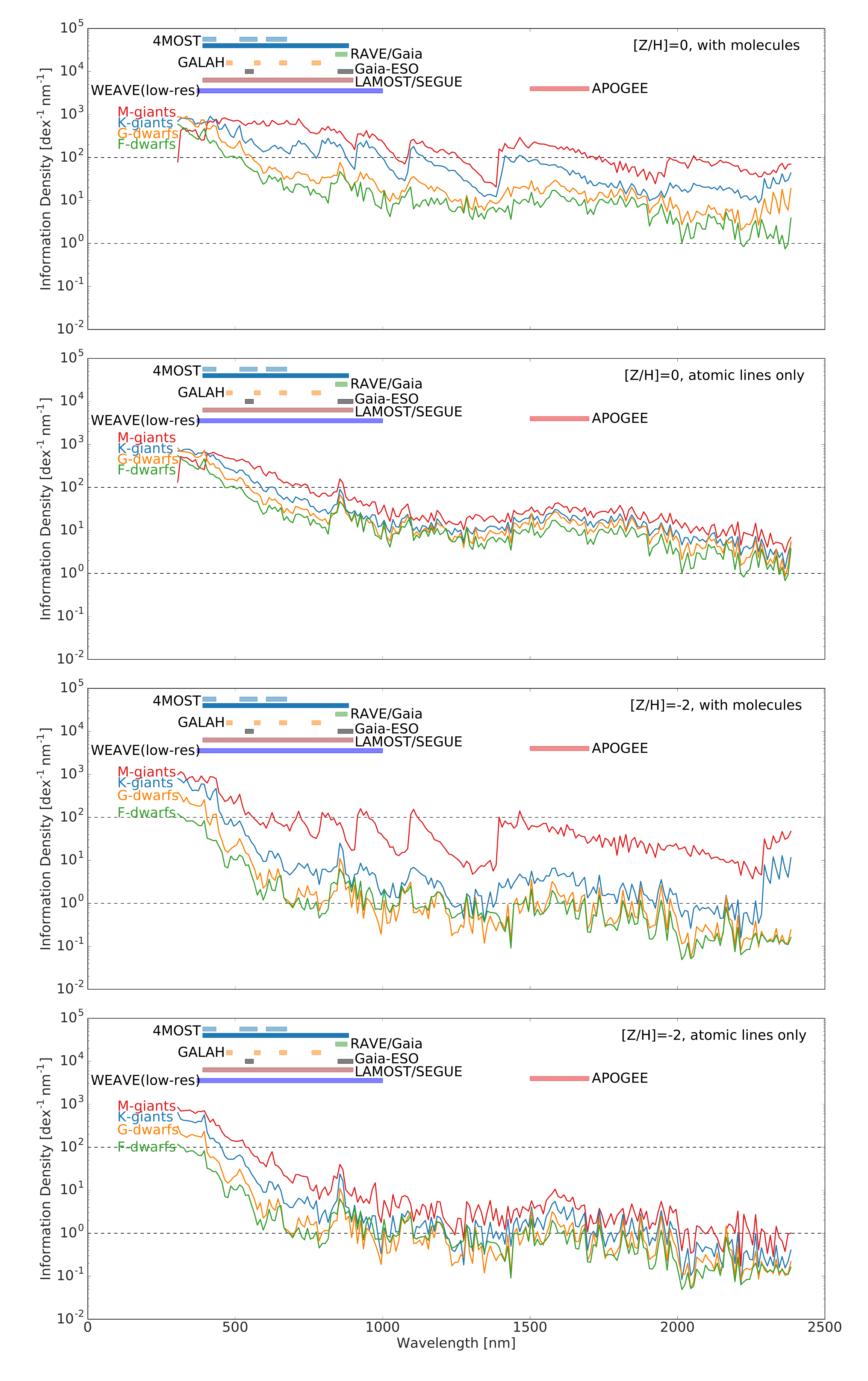}
\caption{Spectral information for all elemental abundances, as a function of wavelength, spectral type (line color) and metallicity (top vs. bottom panels): each line shows the sum of all gradient spectra from elements with atomic numbers from 3 to 99 and take into account that bluer wavelengths have smaller resolution elements. The colored horizontal bars show the wavelength ranges of various large spectroscopic surveys. Lines in different colors illustrate different stellar types -- from M-giants to F-dwarfs. The top two panels show the information content of $[Z/{\rm H}]=0$, and the lower two panels assume $[Z/{\rm H}] = -2$. Within each of two panels at $[Z/{\rm H}] = 0$ and $[Z/{\rm H}] = -2$, the lower one excludes molecular lines in the model. This quantifies that the information content about many elements increases towards shorter wavelengths and cooler spectral types. Also, molecular features, often omitted from analyses because of their complexity, have large information content.}
\label{fig:information-density-1}
\end{figure*}

\begin{figure*}
\centering
\includegraphics[width=0.75\textwidth]{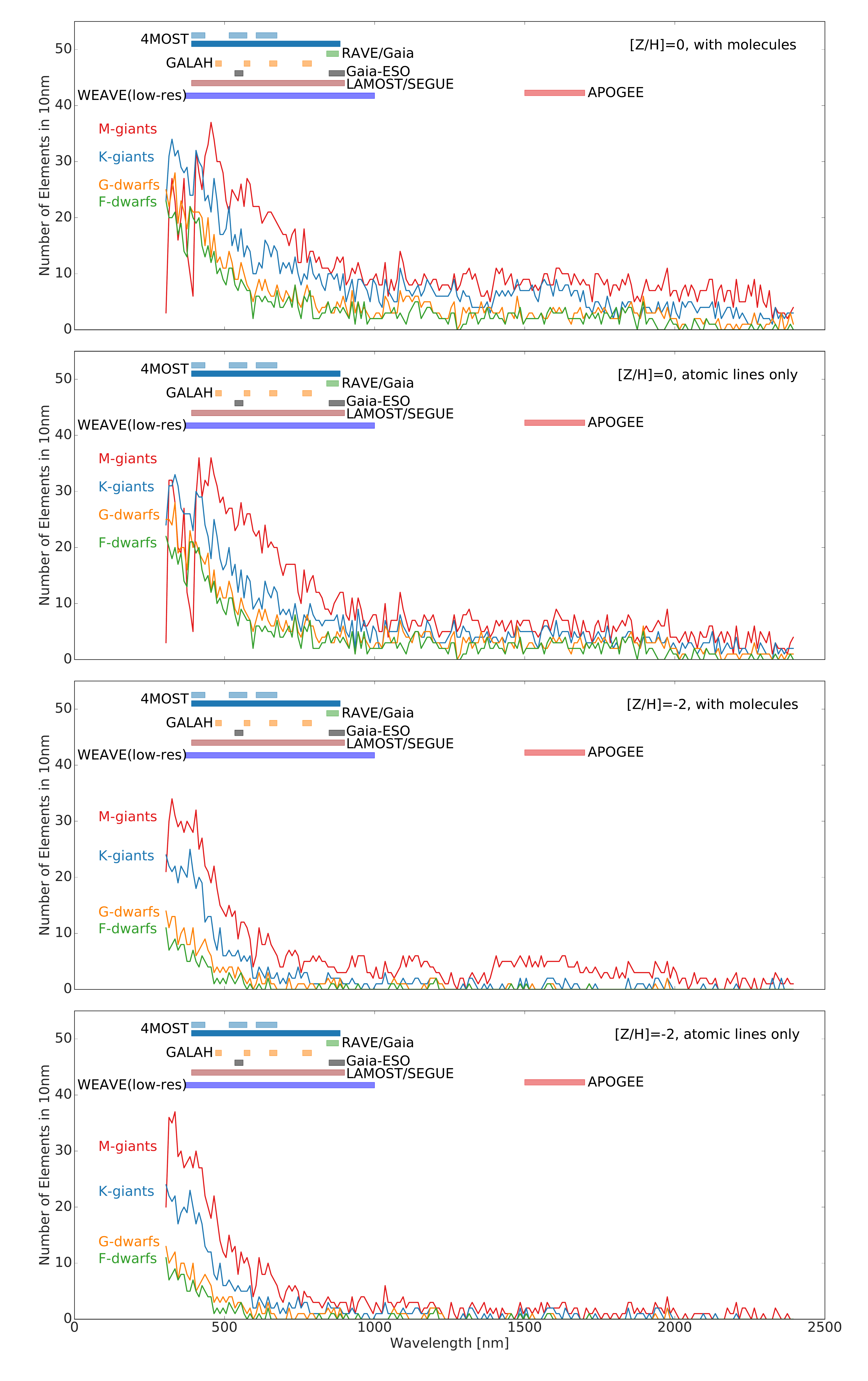}
\caption{Number of elements with detectable spectral signatures within any $10\,$nm portion of the spectrum. The panel layout is the same as Fig.~\ref{fig:information-density-1}. We define an element to have detectable spectral signatures if there is at least a spectral line with gradient greater than $0.025 \, {\rm dex}^{-1}$ at the resolution of $R = \,$6$,$000. Note that the total number of detectable elements (cf. Fig.~\ref{fig:num-elements}) is necessarily larger than the values shown in the $y$-axis because different elements contribute at different wavelengths. Only a few important molecules contribute to most of the information in the infrared. Therefore, despite its high information content as shown in Fig.~\ref{fig:information-density-1}, the number of elements with detectable spectral signatures in the infrared is much smaller than in the optical.}
\label{fig:information-density-2}
\end{figure*}

\begin{figure*}
\centering
\includegraphics[width=1.0\textwidth]{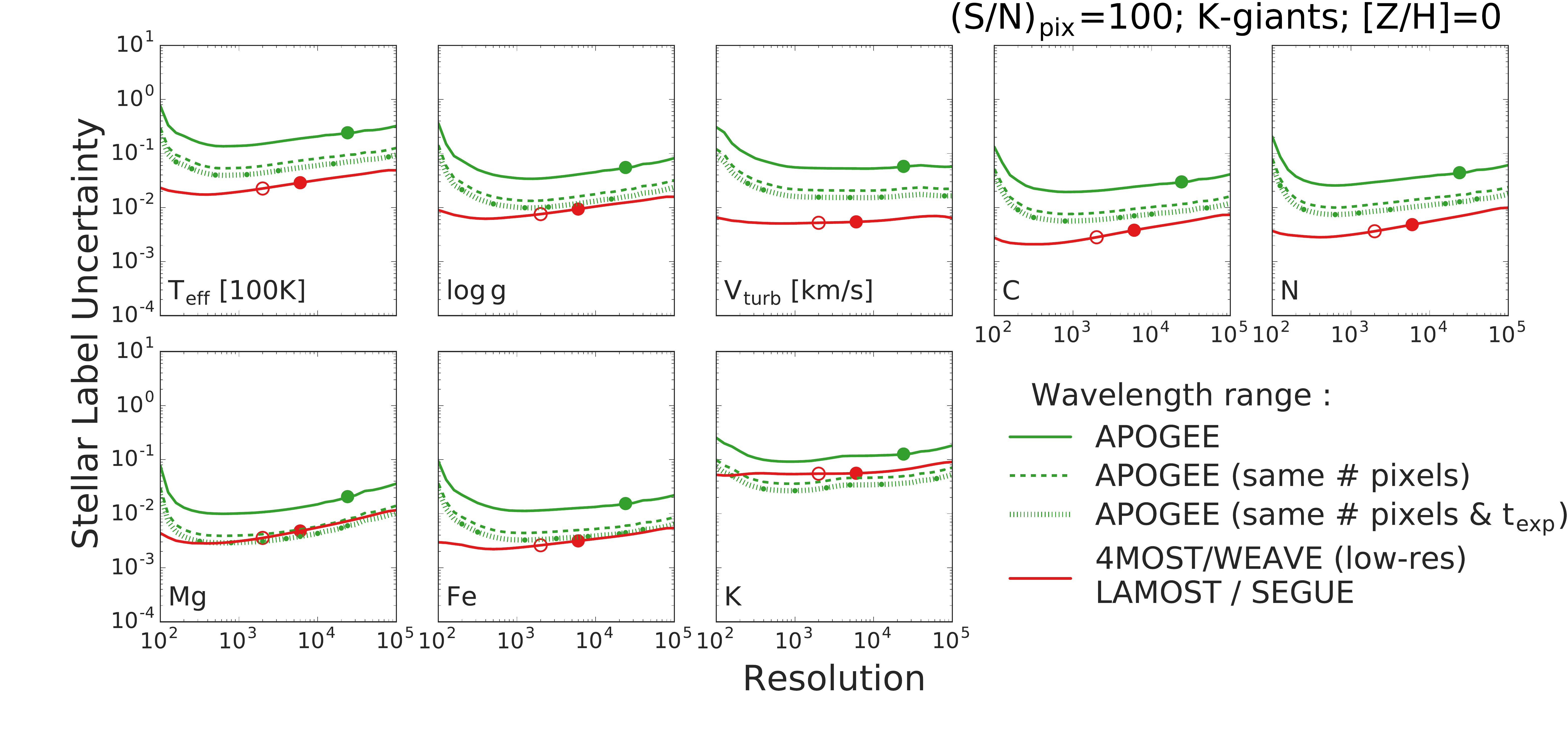}
\caption{Stellar label uncertainties as a function of spectral resolution, assuming the same exposure time and the same number of detector pixels. We assume an anchor point at $R = \,$6$,$000, i.e., we show at $R =\,$6$,$000 the stellar label uncertainties with S/N$\, = 100$ per wavelength pixel and the wavelength range of the spectroscopic surveys. For other spectral resolutions, we vary the S/N according to Poisson statistics assuming the same exposure time (low-resolution spectra have higher S/N) and further scale the label uncertainties by $\sqrt{R/6,000}$, taking into account that low-resolution spectra have a more extensive wavelength range for the same detector real estate. Shown are only a few stellar labels for the purpose of illustration. We find the stellar uncertainties to have a weak dependence on spectral resolution. The solid green lines adopt the wavelength range of APOGEE and the solid red lines adopt the wavelength range of 4MOST/WEAVE (in the low-resolution configuration), SEGUE, and LAMOST. The green symbols, red filled symbols, and red hollow symbols demonstrate the actual survey resolutions of APOGEE, 4MOST/WEAVE (low-resolution) and LAMOST/SEGUE, respectively. Since 4MOST has a larger wavelength range and a smaller resolution element than APOGEE, and red giants are brighter in the infrared than the optical, the green dashed and dotted lines account for these differences in order to have a fairer comparison for the optical and the infrared. The green dashed lines assume that APOGEE has the same number of wavelength pixels as the optical surveys, and the green dotted lines further assume the same exposure time (APOGEE achieves higher S/N because giants are brighter in the infrared).}
\label{fig:no-gain-1}
\end{figure*}

\begin{figure*}
\centering
\includegraphics[width=1.0\textwidth]{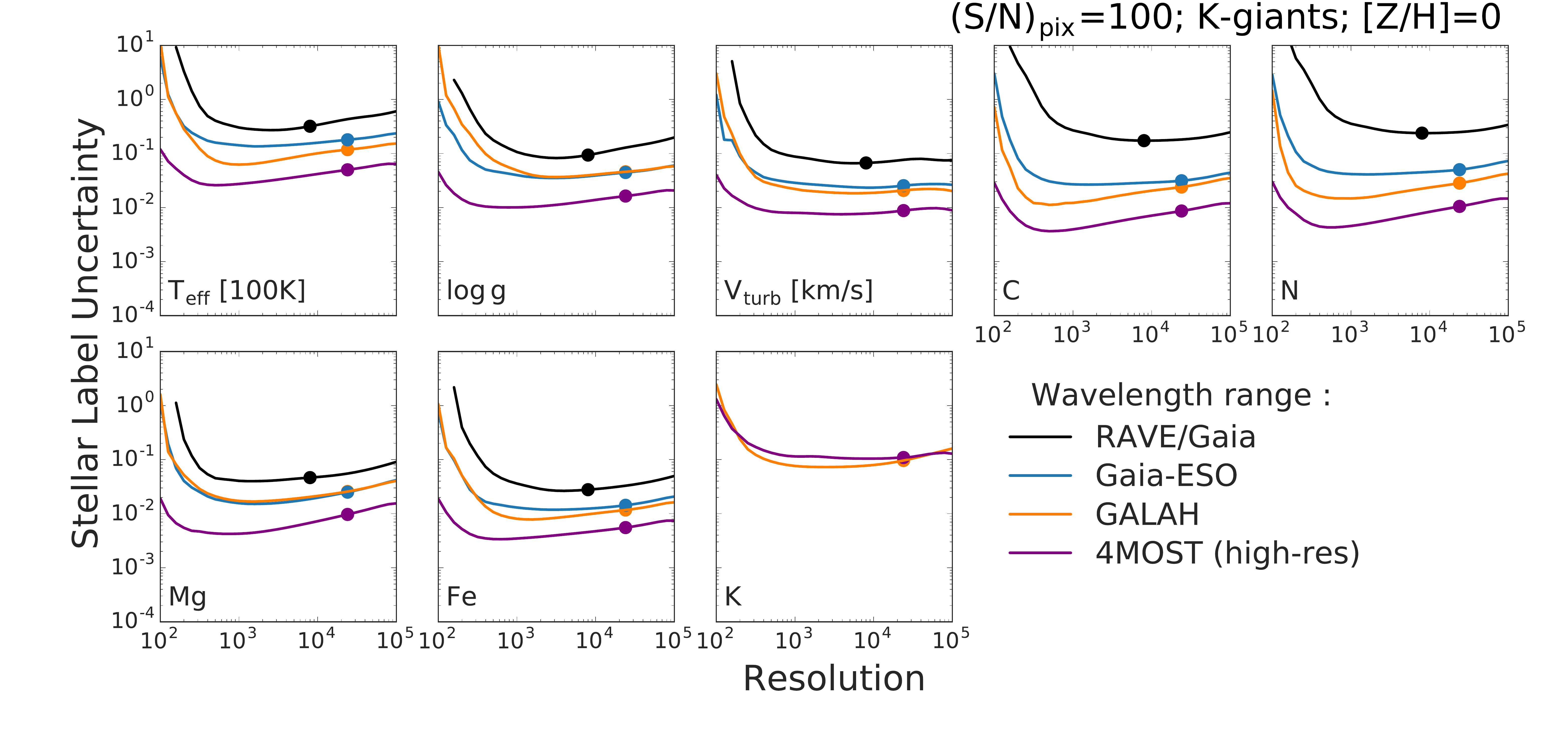}
\caption{Same as Fig.~\ref{fig:no-gain-1}, but for the other spectroscopic surveys under consideration here. Note that RAVE and Gaia RVS do not detect K, so we omit RAVE and Gaia RVS from the potassium subplot.}
\label{fig:no-gain-2}
\end{figure*}

Fig.~\ref{fig:information-density-1} shows the sum of gradient spectra from all elements, illustrating the total spectral information. Since the resolution element is proportional to $\lambda/R$, a bluer wavelength has a smaller resolution element -- in other words, we sample more wavelength pixels at bluer wavelengths. Taking that into account, we further divide the sum of gradients by the wavelength in Fig.~\ref{fig:information-density-1}. Therefore, the $y$-axis has an unit of ${\rm dex}^{-1} {\rm nm}^{-1}$. But it is the relative amount of information that matters, the absolute scale of the $y$-axis is not important. We note that the total information does not directly infer the number of detectable elements. For example, molecules such as TiO and CN can have an enormous amount of spectral lines, but yet there are not many elements involved. To overcome this shortcoming, Fig.~\ref{fig:information-density-2} provides another view of the information content \citep[also see][for a similar analysis]{bla04}. We separate the wavelength range into portions of $10\,$nm, and evaluate how many elements have detectable spectral signatures in each of these portions. We define an element has detectable spectral signatures if there is at least a spectral line with a gradient greater than $0.025 \, {\rm dex}^{-1}$ at $R = \,$6$,$000. The elements that have detectable spectral signatures at each portion can be different, therefore the total number of detectable elements is larger than the value in the $y$-axis. We refer readers to Fig.~\ref{fig:elements-surveys} for the total number of detectable elements of each survey. The horizontal bars in these figures illustrate the wavelength ranges of various spectroscopic surveys. For 4MOST, the long bar shows the wavelength range of the low-resolution configuration, and the split short bars show wavelength range of the high-resolution configuration.

The top two panels show the information content of $[Z/{\rm H}] = 0$, and the bottom two panels show the information content of $[Z/{\rm H}] = -2$. As expected, metal-rich stars contain more information and can detect more elements than metal-poor stars because there are more spectral lines. In each of these two panels, we include the molecular contributions in the top panel and leave them out in the bottom panel. Each panel in Fig.~\ref{fig:information-density-1} and Fig.~\ref{fig:information-density-2} shows a similar monotonous increment of information for cooler stars. The difference in total information content for an M-giant and an F-dwarf can differ up to two orders of magnitude and 20 elements depending on the wavelength. It is not surprising that cooler stars have much more information because many spectral lines form at a lower temperature, especially from molecular contributions in the infrared. However, since most information in the infrared comes from a few important molecules, the number of elements with detectable spectral signatures is much lower for the infrared, despite its high information content. The number of detectable elements per $10\,$nm range is 10--30 in the optical but is fewer than $10$ elements in the infrared. 

Since much of the information in the infrared comes from molecules, and due to the composite nature of molecules, this also vividly demonstrates the importance of methods like the PSM to fit all stellar labels simultaneously. Although optical wavelength contains more information, extinction is much more significant in the optical, therefore, optical surveys are typically limited to the solar neighborhood. Infrared surveys, like APOGEE, are better able to cover a larger region of the Milky Way. Clearly, depending on the science goal, the wavelength range of a spectroscopic survey should be carefully chosen. Interestingly, at optical wavelengths, there are about 10--30 elements per 10$\,$nm range, showing that for surveys that are restricted to small wavelength ranges, such as GALAH, we can still easily measure more than 30 elements. Even for surveys like RAVE or Gaia RVS that have very limited wavelength ranges, the information content suggests that, with robust models, we should be able to detect $\sim 15$ elements. Finally, below $400\,$nm, spectral lines in M-giants become so dense that they form an absorption trough with zero stellar flux. We have virtually zero gradient spectra for most elements in this trough, and as a result, both the information content and the number of elements with detectable spectral signatures drop precipitously for M-giants at wavelengths bluer than $400\,$nm.

\begin{figure*}
\centering
\vspace{1.5cm}
\includegraphics[width=1.0\textwidth]{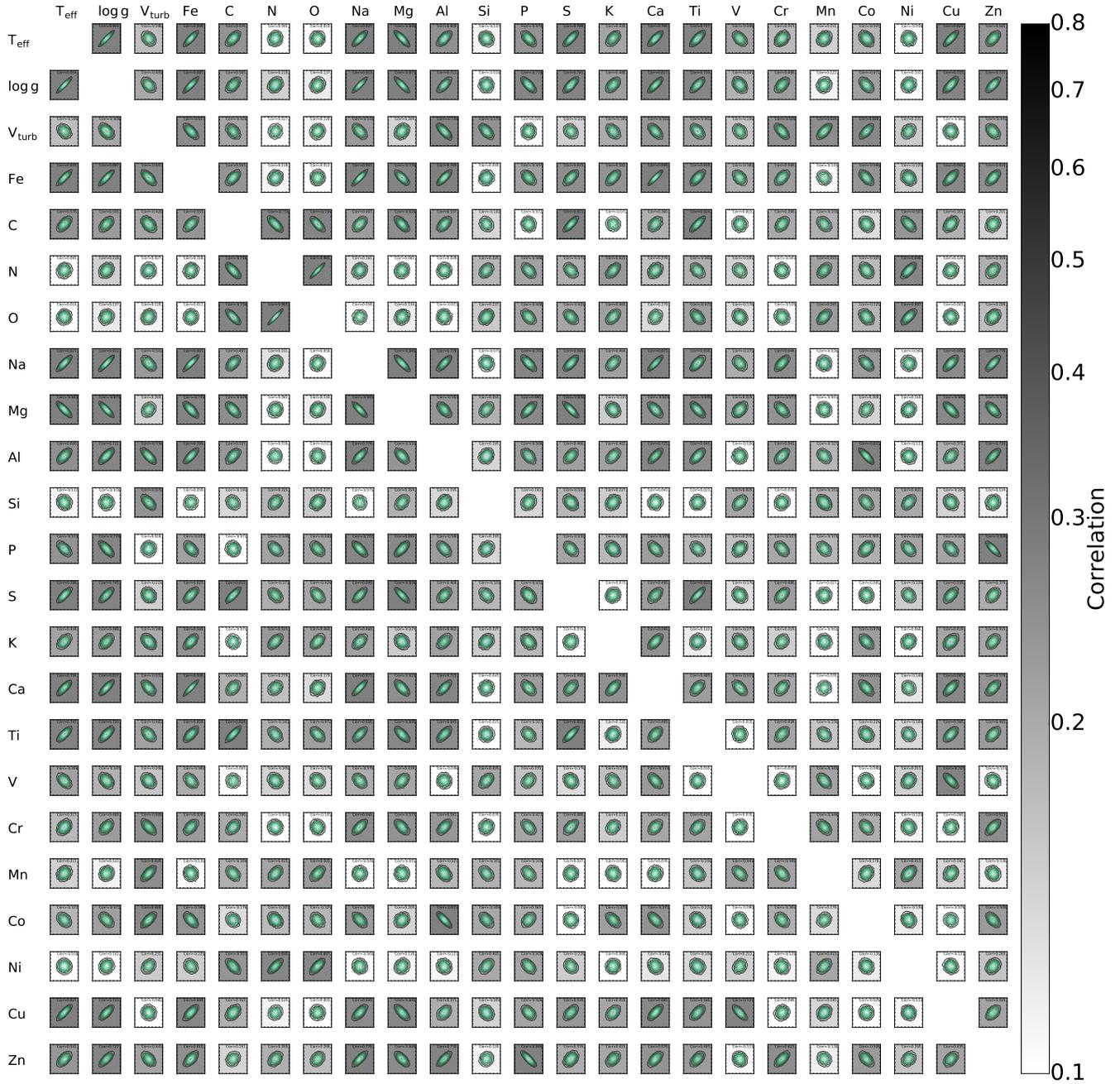}
\caption{Correlations in the estimates of all detectable stellar labels, assuming the wavelength range of APOGEE but at an assumed $R=100$. Each panel shows the correlation of a different label pair, with darker shade indicating a stronger correlation. At $R=100$, most stellar labels are strongly correlated.}
\label{fig:no-correlation-1}
\vspace{5cm}
\end{figure*}

\begin{figure*}
\centering
\vspace{1.5cm}
\includegraphics[width=1.0\textwidth]{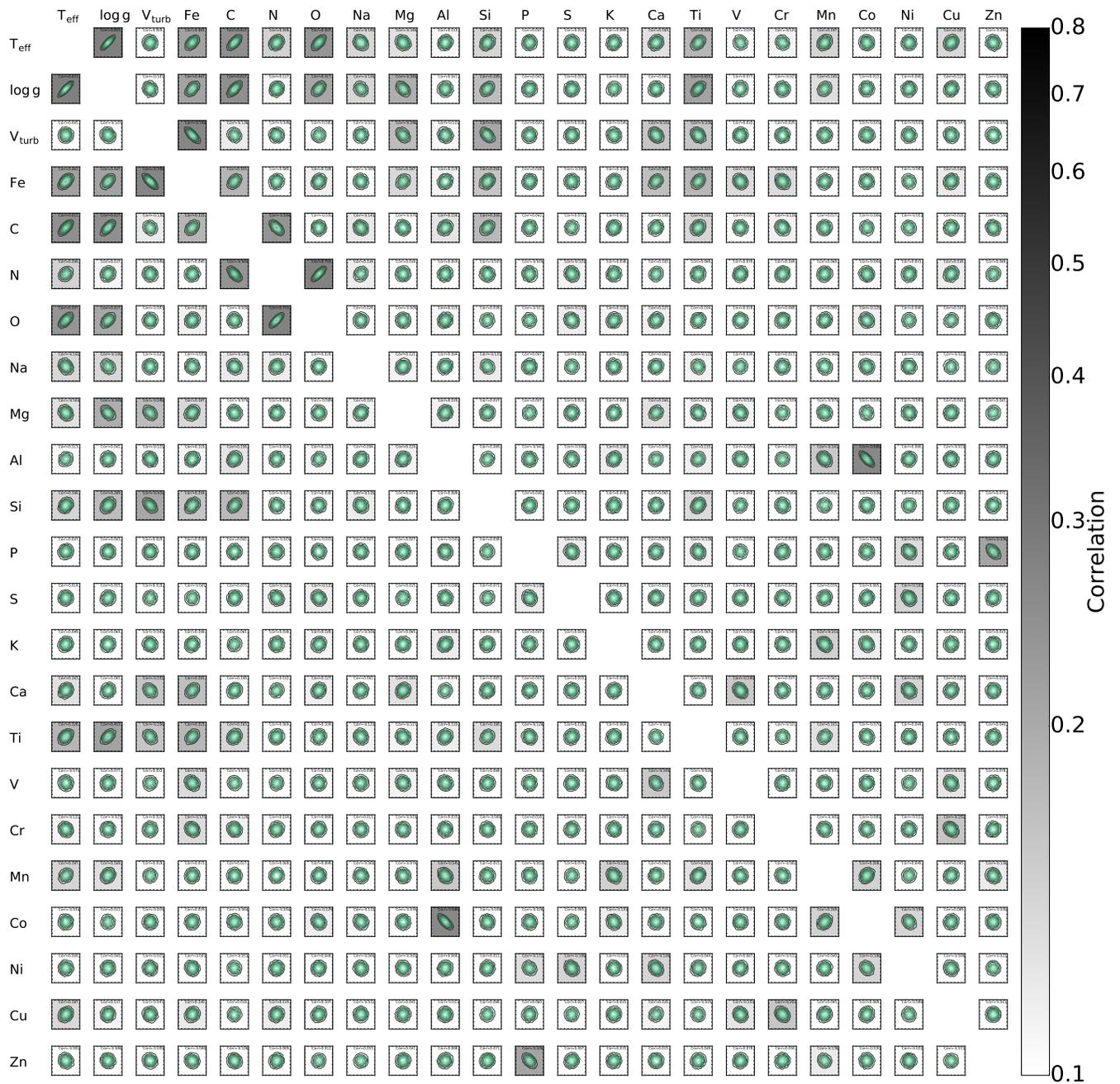}
\caption{Analogous to Fig.~\ref{fig:no-correlation-1}, but for an assumed resolution of $R = \,$1$,$000. At $R = \,$1$,$000, most stellar label estimates are largely uncorrelated. Only those stellar labels that contribute to most wavelength pixels, such as $T_{\rm eff}$, $\log g$, $v_{\rm turb}$, Fe, C, N, O, have strong -- and well-known -- correlations.}
\label{fig:no-correlation-2}
\vspace{5cm}
\end{figure*}

\begin{figure*}
\centering
\vspace{1.5cm}
\includegraphics[width=1.0\textwidth]{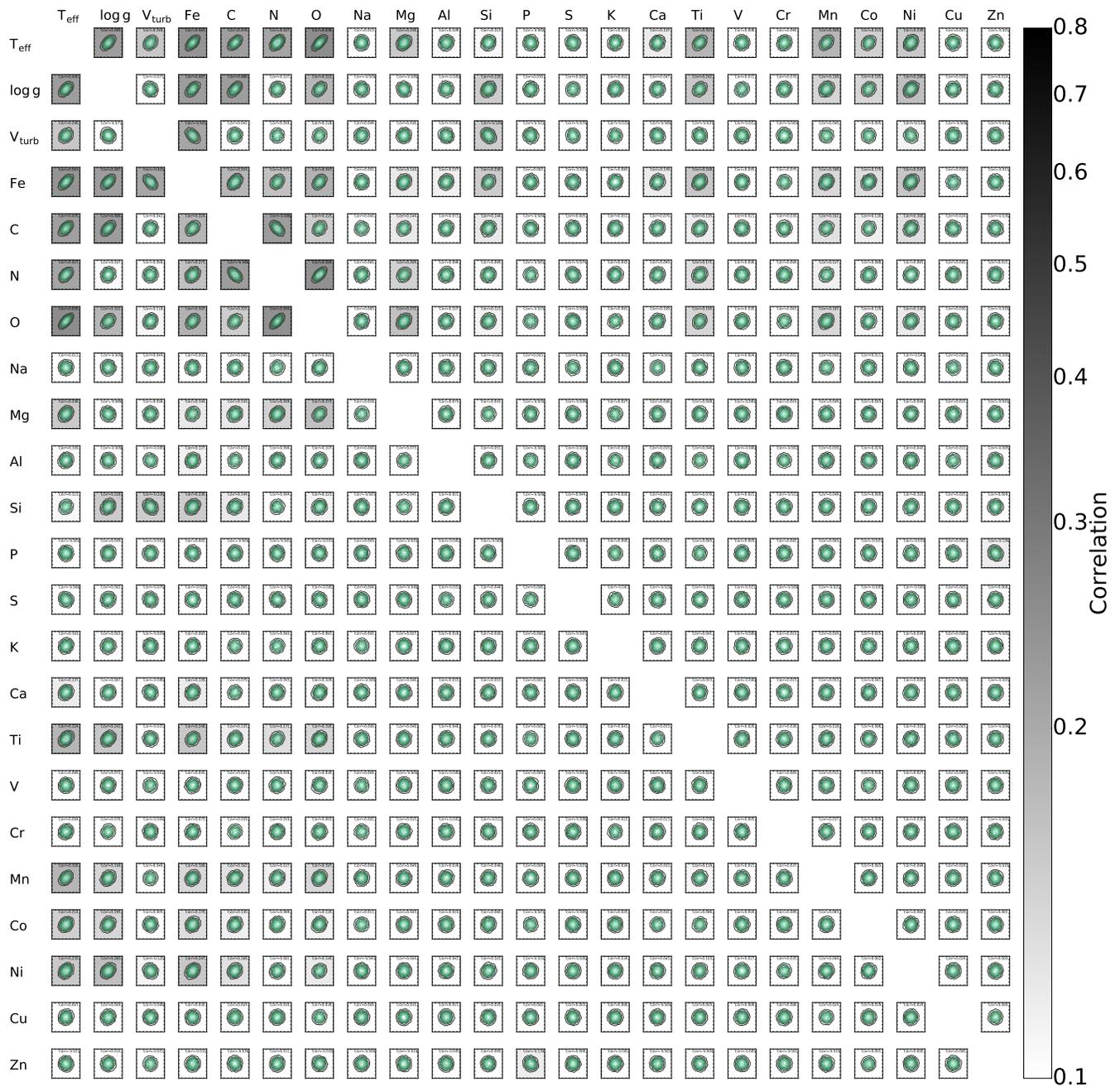}
\caption{Analogous to Fig.~\ref{fig:no-correlation-1}, but here we assume $R = \,$24$,$000. Increasing spectral resolution from $R = \,$1$,$000 to $R = \,$24$,$000 has only minimal effect on the correlations in stellar labels.}
\label{fig:no-correlation-4}
\vspace{5cm}
\end{figure*}

%
%
%
%
%
%

\section{Stellar label uncertainty as a function of spectral resolution}
\label{sec:absolute-uncertainties}

We show in Section~\ref{sec:theoretical-study} and in Fig.~\ref{fig:no-gain-3} that, given the same exposure time and the same number of detector pixels, beyond $R \gtrsim \,$1$,$000, stellar label uncertainties are largely independent of spectral resolution. The gain from a higher S/N and a larger wavelength range for low-resolution spectroscopy compensates the linear trend of uncertainty with resolution when assuming the same S/N and wavelength range. In this appendix, we will study the absolute uncertainties of a few stellar labels to demonstrate this result in more detail. Similar to Section~\ref{sec:theoretical-study}, we assume an anchor point at $R = \,$6$,$000, i.e., at $R = \,$6$,$000, we adopt the wavelength range as assumed and a S/N per wavelength pixel of 100. For other resolutions, we assume a wavelength range inversely proportional to the resolution and a $\sqrt{R/6,000}$ time better/worse in photon noise so that they consume the same number of detector pixels and exposure time. We also assume that spectral information is uniformly distributed over the entire wavelength range, so the larger/smaller wavelength range changes the uncertainty by a factor $\sqrt{R/6,000}$.

Fig.~\ref{fig:no-gain-1} considers the wavelength ranges of the 4MOST survey (optical, $\lambda = 390 -885 \,$nm) and the APOGEE survey (infrared, $\lambda = \,$1$,$500--1$,$700$\,$nm), adopting spectra for solar metallicity K-giants. We only choose a few stellar labels for the purpose of illustration. Although not shown, the other stellar labels follow roughly the same trend. The green and red filled symbols show the survey resolutions of 4MOST (in low-resolution) and APOGEE. Since LAMOST and SEGUE share a similar wavelength range as 4MOST, we overplotted their survey resolutions as red hollow symbols.

Beside the weak dependence of uncertainty with spectral resolution, Fig.~\ref{fig:no-gain-1} also shows that 4MOST has better uncertainties than APOGEE. However, 4MOST also has a larger wavelength range than APOGEE and more wavelength pixels per wavelength range because the resolution element in the bluer wavelength is shorter. Furthermore, red giants are brighter in the infrared than in the optical. For example, the mean flux of a K-giant in the APOGEE wavelength range is about twice of the mean flux in the 4MOST wavelength range. Therefore given the same exposure time, the S/N for the APOGEE survey is about $\sqrt{2}$ better. In order to have a fairer comparison, the green dashed and dotted lines take into account these differences by scaling the APOGEE uncertainties accordingly. Since spectral information adds in quadrature, in the dashed lines, we scale the APOGEE uncertainties in the green solid lines by the square root of the ratio of number of pixels between 4MOST and APOGEE. The dotted lines further scale the uncertainties by a factor of $\sqrt{2}$ due to the brighter flux in the infrared. Fig.~\ref{fig:no-gain-1} shows that even compared to the scaled uncertainties, 4MOST still achieves better precision, demonstrating that the optical wavelength indeed has more spectral information than the infrared, consistent with our assessments in Appendix~\ref{sec:information-contents}. 

Fig.~\ref{fig:no-gain-2} shows similar results, but assuming wavelength ranges from other spectroscopic surveys. Both figures show a weak dependence of uncertainty with resolution beyond $R \gtrsim \,$1$,$000 demonstrating that this trend is generic and is independent of wavelength range. Although not shown in this appendix (cf. Fig.~\ref{fig:no-gain-3}), we also tested that this trend remains the same for other stellar types and metallicities. But on top of these, Fig.~\ref{fig:no-gain-2} also illustrates that surveys having shorter wavelength ranges, e.g., RAVE, Gaia RVS, and GALAH, tend to deviate more from this trend. These surveys have fewer wavelength pixels, as a result, stellar labels become more degenerate and produce deviations from the flat trend seen at higher resolutions.

Throughout this study, we often assume S/N$\,=100$ per pixel. But we emphasize that S/N plays no role in most of our discussions because theoretical uncertainty exactly scales linearly with S/N (cf. Eq.~\ref{eq:CR-calculation}). Since we focuses on the relative uncertainties in this study, the contributions from S/N cancel out. Therefore, our general conclusions in this study are completely independent of S/N.

%
%
%
%
%
%

\section{Correlation of stellar labels as a function of spectral resolution}
\label{sec:correlation-labels}

In Section~\ref{sec:correlation-loss} and Fig.~\ref{fig:no-correlation-5}, we studied the global distribution of correlations from all detectable stellar labels. Here, we show more details in this Appendix. Fig.~\ref{fig:no-correlation-1}--Fig.~\ref{fig:no-correlation-4} show each of the pairwise correlations that comprise the global distribution. We assume the wavelength range of APOGEE, solar metallicity, and K-giants. We define an element to be detectable if its uncertainty at $R = \,$24$,$000 is better than $0.01\,$dex. This definition makes a total 23 stellar labels (20 detectable elements). Fig.~\ref{fig:no-correlation-1}--Fig.~\ref{fig:no-correlation-4} show the pairwise correlations assuming $R = \,$100, 1$,$000 and 24$,$000, respectively -- each panel shows the correlation of a label pair. We shade each panel with the correlation value to guide the eye, and adopt a color scheme in log scale to increase the contrast since most label pairs have moderate correlations between 0.2--0.4. We also tested the other wavelength ranges from different surveys and found that the results remain qualitative the same. 

Fig.~\ref{fig:no-correlation-1} shows that almost all stellar labels are degenerate at $R=100$. At this resolution, there are only $30$ wavelength pixels in the APOGEE wavelength range. Most stellar labels contribute to the same set of pixels. However, when increasing the resolution to $R = \,$1$,$000, as shown in Fig.~\ref{fig:no-correlation-2}, most labels are already not strongly correlated. Only stellar labels that contribute to most pixels -- $T_{\rm eff}$, $\log g$, $v_{\rm turb}$, Fe, C, N, O -- have strong correlations. Going to an even higher resolution, e.g., $R = \,$24$,$000 as shown in Fig.~\ref{fig:no-correlation-4}, no longer decreases the correlations of stellar labels significantly. Stellar labels that have consistent contributions to all pixels continue to correlate in most cases even at the highest resolution. Although not shown, we find that the correlations at $R = \,$100$,$000 remain practically the same as $R = \,$24$,$000. Finally, we note that the correlations evident at high-resolution are often missed by stellar characterization methods that do not solve for all parameters at once, e.g., classical equivalent width based techniques or fitting individual elements with selected windows of strong lines assuming the underlying stellar parameters are fixed.

%
%
%
%
%
%

\end{CJK*}

\vspace{1cm}

\bibliography{biblio.bib}

\begin{thebibliography}{}
\expandafter\ifx\csname natexlab\endcsname\relax\def\natexlab#1{#1}\fi

\bibitem[{{Asplund} {et~al.}(2009){Asplund}, {Grevesse}, {Sauval}, \&
  {Scott}}]{asp09}
{Asplund}, M., {Grevesse}, N., {Sauval}, A.~J., \& {Scott}, P. 2009, \araa, 47,
  481

\bibitem[{{Aumer} {et~al.}(2016){Aumer}, {Binney}, \& {Sch{\"o}nrich}}]{aum16}
{Aumer}, M., {Binney}, J., \& {Sch{\"o}nrich}, R. 2016, ArXiv e-prints,
  arXiv:1607.01972

\bibitem[{{Bensby} {et~al.}(2014){Bensby}, {Feltzing}, \& {Oey}}]{ben14}
{Bensby}, T., {Feltzing}, S., \& {Oey}, M.~S. 2014, \aap, 562, A71

\bibitem[{{Bland-Hawthorn} \& {Freeman}(2004)}]{bla04}
{Bland-Hawthorn}, J., \& {Freeman}, K.~C. 2004, \pasa, 21, 110

\bibitem[{{Bovy}(2016)}]{bov15}
{Bovy}, J. 2016, \apj, 817, 49

\bibitem[{{Brewer} {et~al.}(2015){Brewer}, {Fischer}, {Basu}, {Valenti}, \&
  {Piskunov}}]{bre15}
{Brewer}, J.~M., {Fischer}, D.~A., {Basu}, S., {Valenti}, J.~A., \& {Piskunov},
  N. 2015, \apj, 805, 126

\bibitem[{{Casagrande} {et~al.}(2011){Casagrande}, {Sch{\"o}nrich}, {Asplund},
  {Cassisi}, {Ram{\'{\i}}rez}, {Mel{\'e}ndez}, {Bensby}, \& {Feltzing}}]{cas11}
{Casagrande}, L., {Sch{\"o}nrich}, R., {Asplund}, M., {et~al.} 2011, \aap, 530,
  A138

\bibitem[{{Casey} {et~al.}(2016){Casey}, {Hogg}, {Ness}, {Rix}, {Ho}, \&
  {Gilmore}}]{cas16}
{Casey}, A.~R., {Hogg}, D.~W., {Ness}, M., {et~al.} 2016, ArXiv e-prints,
  arXiv:1603.03040

\bibitem[{{Cramer}(1945)}]{cra45}
{Cramer}, H. 1945, {Mathematical Methods of Statistics} (Princeton University
  Press)

\bibitem[{{Czekala} {et~al.}(2015){Czekala}, {Andrews}, {Mandel}, {Hogg}, \&
  {Green}}]{cze15}
{Czekala}, I., {Andrews}, S.~M., {Mandel}, K.~S., {Hogg}, D.~W., \& {Green},
  G.~M. 2015, \apj, 812, 128

\bibitem[{{De Silva} {et~al.}(2015){De Silva}, {Freeman}, {Bland-Hawthorn},
  {Martell}, {de Boer}, {Asplund}, {Keller}, {Sharma}, {Zucker}, {Zwitter},
  {Anguiano}, {Bacigalupo}, {Bayliss}, {Beavis}, {Bergemann}, {Campbell},
  {Cannon}, {Carollo}, {Casagrande}, {Casey}, {Da Costa}, {D'Orazi}, {Dotter},
  {Duong}, {Heger}, {Ireland}, {Kafle}, {Kos}, {Lattanzio}, {Lewis}, {Lin},
  {Lind}, {Munari}, {Nataf}, {O'Toole}, {Parker}, {Reid}, {Schlesinger},
  {Sheinis}, {Simpson}, {Stello}, {Ting}, {Traven}, {Watson}, {Wittenmyer},
  {Yong}, \& {{\v Z}erjal}}]{des15}
{De Silva}, G.~M., {Freeman}, K.~C., {Bland-Hawthorn}, J., {et~al.} 2015,
  \mnras, 449, 2604

\bibitem[{{Fischer} \& {Valenti}(2005)}]{fis05}
{Fischer}, D.~A., \& {Valenti}, J. 2005, \apj, 622, 1102

\bibitem[{{Fran{\c c}ois} {et~al.}(2016){Fran{\c c}ois}, {Monaco}, {Bonifacio},
  {Moni Bidin}, {Geisler}, \& {Sbordone}}]{fra16}
{Fran{\c c}ois}, P., {Monaco}, L., {Bonifacio}, P., {et~al.} 2016, \aap, 588,
  A7

\bibitem[{{Freeman} \& {Bland-Hawthorn}(2002)}]{fre02}
{Freeman}, K., \& {Bland-Hawthorn}, J. 2002, \araa, 40, 487

\bibitem[{{Garc{\'{\i}}a P{\'e}rez} {et~al.}(2016){Garc{\'{\i}}a P{\'e}rez},
  {Allende Prieto}, {Holtzman}, {Shetrone}, {M{\'e}sz{\'a}ros}, {Bizyaev},
  {Carrera}, {Cunha}, {Garc{\'{\i}}a-Hern{\'a}ndez}, {Johnson}, {Majewski},
  {Nidever}, {Schiavon}, {Shane}, {Smith}, {Sobeck}, {Troup}, {Zamora}, {Bovy},
  {Eisenstein}, {Feuillet}, {Frinchaboy}, {Hayden}, {Hearty}, {Nguyen},
  {O'Connell}, {Pinsonneault}, {Weinberg}, {Wilson}, \& {Zasowski}}]{gar16}
{Garc{\'{\i}}a P{\'e}rez}, A.~E., {Allende Prieto}, C., {Holtzman}, J.~A.,
  {et~al.} 2016, \aj, 151, 144

\bibitem[{{Grand} {et~al.}(2015){Grand}, {Kawata}, \& {Cropper}}]{gra15}
{Grand}, R.~J.~J., {Kawata}, D., \& {Cropper}, M. 2015, \mnras, 447, 4018

\bibitem[{{Grand} {et~al.}(2016){Grand}, {Springel}, {G{\'o}mez}, {Marinacci},
  {Pakmor}, {Campbell}, \& {Jenkins}}]{gra16}
{Grand}, R.~J.~J., {Springel}, V., {G{\'o}mez}, F.~A., {et~al.} 2016, \mnras,
  459, 199

\bibitem[{{Gustafsson} {et~al.}(2008){Gustafsson}, {Edvardsson}, {Eriksson},
  {J{\o}rgensen}, {Nordlund}, \& {Plez}}]{gus08}
{Gustafsson}, B., {Edvardsson}, B., {Eriksson}, K., {et~al.} 2008, \aap, 486,
  951

\bibitem[{{Hauschildt} {et~al.}(1999){Hauschildt}, {Allard}, {Ferguson},
  {Baron}, \& {Alexander}}]{hau99}
{Hauschildt}, P.~H., {Allard}, F., {Ferguson}, J., {Baron}, E., \& {Alexander},
  D.~R. 1999, \apj, 525, 871

\bibitem[{{Hawkins} {et~al.}(2015){Hawkins}, {Jofr{\'e}}, {Masseron}, \&
  {Gilmore}}]{haw15}
{Hawkins}, K., {Jofr{\'e}}, P., {Masseron}, T., \& {Gilmore}, G. 2015, \mnras,
  453, 758

\bibitem[{{Hayden} {et~al.}(2015){Hayden}, {Bovy}, {Holtzman}, {Nidever},
  {Bird}, {Weinberg}, {Andrews}, {Majewski}, {Allende Prieto}, {Anders},
  {Beers}, {Bizyaev}, {Chiappini}, {Cunha}, {Frinchaboy},
  {Garc{\'{\i}}a-Her{\'n}andez}, {Garc{\'{\i}}a P{\'e}rez}, {Girardi},
  {Harding}, {Hearty}, {Johnson}, {M{\'e}sz{\'a}ros}, {Minchev}, {O'Connell},
  {Pan}, {Robin}, {Schiavon}, {Schneider}, {Schultheis}, {Shetrone},
  {Skrutskie}, {Steinmetz}, {Smith}, {Wilson}, {Zamora}, \& {Zasowski}}]{hay15}
{Hayden}, M.~R., {Bovy}, J., {Holtzman}, J.~A., {et~al.} 2015, \apj, 808, 132

\bibitem[{{Heiter} {et~al.}(2015){Heiter}, {Jofr{\'e}}, {Gustafsson}, {Korn},
  {Soubiran}, \& {Th{\'e}venin}}]{hei15}
{Heiter}, U., {Jofr{\'e}}, P., {Gustafsson}, B., {et~al.} 2015, \aap, 582, A49

\bibitem[{{Ho} {et~al.}(2016){Ho}, {Rix}, {Ness}, {Hogg}, {Liu}, \&
  {Ting}}]{ho16}
{Ho}, A.~Y.~Q., {Rix}, H.-W., {Ness}, M.~K., {et~al.} 2016, ArXiv e-prints,
  arXiv:1609.03195

\bibitem[{{Ho} {et~al.}(2017){Ho}, {Ness}, {Hogg}, {Rix}, {Liu}, {Yang},
  {Zhang}, {Hou}, \& {Wang}}]{ho17}
{Ho}, A.~Y.~Q., {Ness}, M.~K., {Hogg}, D.~W., {et~al.} 2017, \apj, 836, 5

\bibitem[{{Hogg} {et~al.}(2016){Hogg}, {Casey}, {Ness}, {Rix}, \&
  {Foreman-Mackey}}]{hog16}
{Hogg}, D.~W., {Casey}, A.~R., {Ness}, M., {Rix}, H.-W., \& {Foreman-Mackey},
  D. 2016, ArXiv e-prints, arXiv:1601.05413

\bibitem[{{Holtzman} {et~al.}(2015){Holtzman}, {Shetrone}, {Johnson}, {Allende
  Prieto}, {Anders}, {Andrews}, {Beers}, {Bizyaev}, {Blanton}, {Bovy},
  {Carrera}, {Chojnowski}, {Cunha}, {Eisenstein}, {Feuillet}, {Frinchaboy},
  {Galbraith-Frew}, {Garc{\'{\i}}a P{\'e}rez}, {Garc{\'{\i}}a-Hern{\'a}ndez},
  {Hasselquist}, {Hayden}, {Hearty}, {Ivans}, {Majewski}, {Martell},
  {Meszaros}, {Muna}, {Nidever}, {Nguyen}, {O'Connell}, {Pan}, {Pinsonneault},
  {Robin}, {Schiavon}, {Shane}, {Sobeck}, {Smith}, {Troup}, {Weinberg},
  {Wilson}, {Wood-Vasey}, {Zamora}, \& {Zasowski}}]{hol15}
{Holtzman}, J.~A., {Shetrone}, M., {Johnson}, J.~A., {et~al.} 2015, \aj, 150,
  148

\bibitem[{{Iwamoto} {et~al.}(1999){Iwamoto}, {Brachwitz}, {Nomoto},
  {Kishimoto}, {Umeda}, {Hix}, \& {Thielemann}}]{iwa99}
{Iwamoto}, K., {Brachwitz}, F., {Nomoto}, K., {et~al.} 1999, \apjs, 125, 439

\bibitem[{{Jofr{\'e}} {et~al.}(2014){Jofr{\'e}}, {Heiter}, {Soubiran},
  {Blanco-Cuaresma}, {Worley}, {Pancino}, {Cantat-Gaudin}, {Magrini},
  {Bergemann}, {Gonz{\'a}lez Hern{\'a}ndez}, {Hill}, {Lardo}, {de Laverny},
  {Lind}, {Masseron}, {Montes}, {Mucciarelli}, {Nordlander}, {Recio Blanco},
  {Sobeck}, {Sordo}, {Sousa}, {Tabernero}, {Vallenari}, \& {Van Eck}}]{jof14}
{Jofr{\'e}}, P., {Heiter}, U., {Soubiran}, C., {et~al.} 2014, \aap, 564, A133

\bibitem[{{Jofr{\'e}} {et~al.}(2015){Jofr{\'e}}, {Heiter}, {Soubiran},
  {Blanco-Cuaresma}, {Masseron}, {Nordlander}, {Chemin}, {Worley}, {Van Eck},
  {Hourihane}, {Gilmore}, {Adibekyan}, {Bergemann}, {Cantat-Gaudin},
  {Delgado-Mena}, {Gonz{\'a}lez Hern{\'a}ndez}, {Guiglion}, {Lardo}, {de
  Laverny}, {Lind}, {Magrini}, {Mikolaitis}, {Montes}, {Pancino},
  {Recio-Blanco}, {Sordo}, {Sousa}, {Tabernero}, \& {Vallenari}}]{jof15}
---. 2015, \aap, 582, A81

\bibitem[{{Karakas} \& {Lattanzio}(2014)}]{kar14}
{Karakas}, A.~I., \& {Lattanzio}, J.~C. 2014, \pasa, 31, 30

\bibitem[{{Kawata} {et~al.}(2016){Kawata}, {Grand}, {Gibson}, {Casagrande},
  {Hunt}, \& {Brook}}]{kaw16}
{Kawata}, D., {Grand}, R.~J.~J., {Gibson}, B.~K., {et~al.} 2016, ArXiv
  e-prints, arXiv:1604.07412

\bibitem[{{Kirby} {et~al.}(2010){Kirby}, {Guhathakurta}, {Simon}, {Geha},
  {Rockosi}, {Sneden}, {Cohen}, {Sohn}, {Majewski}, \& {Siegel}}]{kir10}
{Kirby}, E.~N., {Guhathakurta}, P., {Simon}, J.~D., {et~al.} 2010, \apjs, 191,
  352

\bibitem[{{Kurucz}(1970)}]{kur70}
{Kurucz}, R.~L. 1970, SAO Special Report, 309

\bibitem[{{Kurucz}(1993)}]{kur93}
---. 1993, {SYNTHE spectrum synthesis programs and line data}, ed. {Kurucz,
  R.~L.}

\bibitem[{{Kurucz}(1996)}]{kur96}
{Kurucz}, R.~L. 1996, in Astronomical Society of the Pacific Conference Series,
  Vol. 108, M.A.S.S., Model Atmospheres and Spectrum Synthesis, ed. S.~J.
  {Adelman}, F.~{Kupka}, \& W.~W. {Weiss}, 2

\bibitem[{{Kurucz}(2003)}]{kur03}
{Kurucz}, R.~L. 2003, in IAU Symposium, Vol. 210, Modelling of Stellar
  Atmospheres, ed. N.~{Piskunov}, W.~W. {Weiss}, \& D.~F. {Gray}, 45

\bibitem[{{Kurucz}(2005)}]{kur05}
---. 2005, Memorie della Societa Astronomica Italiana Supplementi, 8, 14

\bibitem[{{Kurucz} \& {Avrett}(1981)}]{kur81}
{Kurucz}, R.~L., \& {Avrett}, E.~H. 1981, SAO Special Report, 391

\bibitem[{{Lee} {et~al.}(2011){Lee}, {Beers}, {Allende Prieto}, {Lai},
  {Rockosi}, {Morrison}, {Johnson}, {An}, {Sivarani}, \& {Yanny}}]{lee11}
{Lee}, Y.~S., {Beers}, T.~C., {Allende Prieto}, C., {et~al.} 2011, \aj, 141, 90

\bibitem[{{Lee} {et~al.}(2013){Lee}, {Beers}, {Masseron}, {Plez}, {Rockosi},
  {Sobeck}, {Yanny}, {Lucatello}, {Sivarani}, {Placco}, \& {Carollo}}]{lee13}
{Lee}, Y.~S., {Beers}, T.~C., {Masseron}, T., {et~al.} 2013, \aj, 146, 132

\bibitem[{{Limongi} \& {Chieffi}(2003)}]{lim03}
{Limongi}, M., \& {Chieffi}, A. 2003, \apj, 592, 404

\bibitem[{{Lind} {et~al.}(2013){Lind}, {Melendez}, {Asplund}, {Collet}, \&
  {Magic}}]{lind13}
{Lind}, K., {Melendez}, J., {Asplund}, M., {Collet}, R., \& {Magic}, Z. 2013,
  \aap, 554, A96

\bibitem[{{Lindegren} \& {Feltzing}(2013)}]{lin13}
{Lindegren}, L., \& {Feltzing}, S. 2013, \aap, 553, A94

\bibitem[{{Liu} {et~al.}(2016){Liu}, {Yong}, {Asplund}, {Ram{\'{\i}}rez}, \&
  {Mel{\'e}ndez}}]{liu16}
{Liu}, F., {Yong}, D., {Asplund}, M., {Ram{\'{\i}}rez}, I., \& {Mel{\'e}ndez},
  J. 2016, \mnras, 457, 3934

\bibitem[{{Luo} {et~al.}(2015){Luo}, {Zhao}, {Zhao}, {Deng}, {Liu}, {Jing},
  {Wang}, {Zhang}, {Shi}, {Cui}, {Chu}, {Li}, {Bai}, {Wu}, {Cai}, {Cao}, {Cao},
  {Carlin}, {Chen}, {Chen}, {Chen}, {Chen}, {Chen}, {Chen}, {Chen},
  {Christlieb}, {Chu}, {Cui}, {Dong}, {Du}, {Fan}, {Feng}, {Fu}, {Gao}, {Gong},
  {Gu}, {Guo}, {Han}, {He}, {Hou}, {Hou}, {Hou}, {Hu}, {Hu}, {Hu}, {Huo},
  {Jia}, {Jiang}, {Jiang}, {Jiang}, {Jin}, {Kong}, {Kong}, {Lei}, {Li}, {Li},
  {Li}, {Li}, {Li}, {Li}, {Li}, {Li}, {Li}, {Li}, {Li}, {Li}, {Liang}, {Lin},
  {Liu}, {Liu}, {Liu}, {Liu}, {Lu}, {Luo}, {Mao}, {Newberg}, {Ni}, {Qi}, {Qi},
  {Shen}, {Shi}, {Song}, {Song}, {Su}, {Su}, {Tang}, {Tao}, {Tian}, {Wang},
  {Wang}, {Wang}, {Wang}, {Wang}, {Wang}, {Wang}, {Wang}, {Wang}, {Wang},
  {Wang}, {Wang}, {Wang}, {Wang}, {Wang}, {Wang}, {Wang}, {Wang}, {Wang},
  {Wang}, {Wei}, {Wei}, {Wu}, {Wu}, {Wu}, {Wu}, {Xing}, {Xu}, {Xu}, {Xu},
  {Yan}, {Yang}, {Yang}, {Yang}, {Yang}, {Yao}, {Yu}, {Yuan}, {Yuan}, {Yuan},
  {Yuan}, {Zhai}, {Zhang}, {Zhang}, {Zhang}, {Zhang}, {Zhang}, {Zhang},
  {Zhang}, {Zhang}, {Zhao}, {Zhou}, {Zhou}, {Zhu}, {Zhu}, {Zou}, \&
  {Zuo}}]{luo15}
{Luo}, A.-L., {Zhao}, Y.-H., {Zhao}, G., {et~al.} 2015, Research in Astronomy
  and Astrophysics, 15, 1095

\bibitem[{{Majewski} {et~al.}(2015){Majewski}, {Schiavon}, {Frinchaboy},
  {Allende Prieto}, {Barkhouser}, {Bizyaev}, {Blank}, {Brunner}, {Burton},
  {Carrera}, {Chojnowski}, {Cunha}, {Epstein}, {Fitzgerald}, {Garcia Perez},
  {Hearty}, {Henderson}, {Holtzman}, {Johnson}, {Lam}, {Lawler}, {Maseman},
  {Meszaros}, {Nelson}, {Coung Nguyen}, {Nidever}, {Pinsonneault}, {Shetrone},
  {Smee}, {Smith}, {Stolberg}, {Skrutskie}, {Walker}, {Wilson}, {Zasowski},
  {Anders}, {Basu}, {Beland}, {Blanton}, {Bovy}, {Brownstein}, {Carlberg},
  {Chaplin}, {Chiappini}, {Eisenstein}, {Elsworth}, {Feuillet}, {Fleming},
  {Galbraith-Frew}, {Garcia}, {Anibal Garcia-Hernandez}, {Gillespie},
  {Girardi}, {Gunn}, {Hasselquist}, {Hayden}, {Hekker}, {Ivans}, {Kinemuchi},
  {Klaene}, {Mahadevan}, {Mathur}, {Mosser}, {Muna}, {Munn}, {Nichol},
  {O'Connell}, {Robin}, {Rocha-Pinto}, {Schultheis}, {Serenelli}, {Shane},
  {Silva Aguirre}, {Sobeck}, {Thompson}, {Troup}, {Weinberg}, \&
  {Zamora}}]{maj15}
{Majewski}, S.~R., {Schiavon}, R.~P., {Frinchaboy}, P.~M., {et~al.} 2015, ArXiv
  e-prints, arXiv:1509.05420

\bibitem[{{Martell} \& {Grebel}(2010)}]{mar10}
{Martell}, S.~L., \& {Grebel}, E.~K. 2010, \aap, 519, A14

\bibitem[{{Martell} {et~al.}(2016){Martell}, {Shetrone}, {Lucatello},
  {Schiavon}, {M{\'e}sz{\'a}ros}, {Allende Prieto}, {Garc{\'{\i}}a
  Hern{\'a}ndez}, {Beers}, \& {Nidever}}]{mar16}
{Martell}, S.~L., {Shetrone}, M.~D., {Lucatello}, S., {et~al.} 2016, \apj, 825,
  146

\bibitem[{{Martig} {et~al.}(2014){Martig}, {Minchev}, \& {Flynn}}]{mar14}
{Martig}, M., {Minchev}, I., \& {Flynn}, C. 2014, \mnras, 443, 2452

\bibitem[{{Minchev} {et~al.}(2014){Minchev}, {Chiappini}, \& {Martig}}]{min14}
{Minchev}, I., {Chiappini}, C., \& {Martig}, M. 2014, \aap, 572, A92

\bibitem[{{Ness} {et~al.}(2015){Ness}, {Hogg}, {Rix}, {Ho}, \&
  {Zasowski}}]{nes15a}
{Ness}, M., {Hogg}, D.~W., {Rix}, H.-W., {Ho}, A.~Y.~Q., \& {Zasowski}, G.
  2015, \apj, 808, 16

\bibitem[{{Nomoto} {et~al.}(2013){Nomoto}, {Kobayashi}, \& {Tominaga}}]{nom13}
{Nomoto}, K., {Kobayashi}, C., \& {Tominaga}, N. 2013, \araa, 51, 457

\bibitem[{{Quillen} {et~al.}(2015){Quillen}, {Anguiano}, {De Silva}, {Freeman},
  {Zucker}, {Minchev}, \& {Bland-Hawthorn}}]{qui15}
{Quillen}, A.~C., {Anguiano}, B., {De Silva}, G., {et~al.} 2015, \mnras, 450,
  2354

\bibitem[{{Rao}(1945)}]{rao45}
{Rao}, C.~R. 1945, Bulletin of the Calcutta Mathematical Society, 37, 81

\bibitem[{{Recio-Blanco} {et~al.}(2016){Recio-Blanco}, {de Laverny}, {Allende
  Prieto}, {Fustes}, {Manteiga}, {Arcay}, {Bijaoui}, {Dafonte}, {Ordenovic}, \&
  {Ordo{\~n}ez Blanco}}]{rec16}
{Recio-Blanco}, A., {de Laverny}, P., {Allende Prieto}, C., {et~al.} 2016,
  \aap, 585, A93

\bibitem[{{Rix} \& {Bovy}(2013)}]{rix13}
{Rix}, H.-W., \& {Bovy}, J. 2013, \aapr, 21, 61

\bibitem[{{Rix} {et~al.}(2016){Rix}, {Ting}, {Conroy}, \& {Hogg}}]{rix16}
{Rix}, H.-W., {Ting}, Y.-S., {Conroy}, C., \& {Hogg}, D.~W. 2016, \apjl, 826,
  L25

\bibitem[{{Schiavon} {et~al.}(2016){Schiavon}, {Zamora}, {Carrera},
  {Lucatello}, {Robin}, {Ness}, {Martell}, {Smith}, {Garcia Hernandez},
  {Manchado}, {Schoenrich}, {Bastian}, {Chiappini}, {Shetrone}, {Mackereth},
  {Williams}, {Meszaros}, {Allende Prieto}, {Anders}, {Bizyaev}, {Beers},
  {Chojnowski}, {Cunha}, {Epstein}, {Frinchaboy}, {Garcia Perez}, {Hearty},
  {Holtzman}, {Johnson}, {Kinemuchi}, {Majewski}, {Muna}, {Nidever}, {Nguyen},
  {O'Connell}, {Oravetz}, {Pan}, {Pinsonneault}, {Schneider}, {Schultheis},
  {Simmons}, {Skrutskie}, {Sobeck}, {Wilson}, \& {Zasowski}}]{schi16}
{Schiavon}, R.~P., {Zamora}, O., {Carrera}, R., {et~al.} 2016, ArXiv e-prints,
  arXiv:1606.05651

\bibitem[{{Sch{\"o}nrich} \& {Binney}(2009)}]{sch09}
{Sch{\"o}nrich}, R., \& {Binney}, J. 2009, \mnras, 396, 203

\bibitem[{{Sch{\"o}nrich} \& {McMillan}(2016)}]{sch16}
{Sch{\"o}nrich}, R., \& {McMillan}, P. 2016, ArXiv e-prints, arXiv:1605.02338

\bibitem[{{SDSS Collaboration} {et~al.}(2016){SDSS Collaboration}, {Albareti},
  {Allende Prieto}, {Almeida}, {Anders}, {Anderson}, {Andrews},
  {Aragon-Salamanca}, {Argudo-Fernandez}, {Armengaud}, \& et~al.}]{sds16}
{SDSS Collaboration}, {Albareti}, F.~D., {Allende Prieto}, C., {et~al.} 2016,
  ArXiv e-prints, arXiv:1608.02013

\bibitem[{{Smiljanic} {et~al.}(2014){Smiljanic}, {Korn}, {Bergemann}, {Frasca},
  {Magrini}, {Masseron}, {Pancino}, {Ruchti}, {San Roman}, {Sbordone}, {Sousa},
  {Tabernero}, {Tautvai{\v s}ien{\.e}}, {Valentini}, {Weber}, {Worley},
  {Adibekyan}, {Allende Prieto}, {Barisevi{\v c}ius}, {Biazzo},
  {Blanco-Cuaresma}, {Bonifacio}, {Bragaglia}, {Caffau}, {Cantat-Gaudin},
  {Chorniy}, {de Laverny}, {Delgado-Mena}, {Donati}, {Duffau}, {Franciosini},
  {Friel}, {Geisler}, {Gonz{\'a}lez Hern{\'a}ndez}, {Gruyters}, {Guiglion},
  {Hansen}, {Heiter}, {Hill}, {Jacobson}, {Jofre}, {J{\"o}nsson}, {Lanzafame},
  {Lardo}, {Ludwig}, {Maiorca}, {Mikolaitis}, {Montes}, {Morel}, {Mucciarelli},
  {Mu{\~n}oz}, {Nordlander}, {Pasquini}, {Puzeras}, {Recio-Blanco}, {Ryde},
  {Sacco}, {Santos}, {Serenelli}, {Sordo}, {Soubiran}, {Spina}, {Steffen},
  {Vallenari}, {Van Eck}, {Villanova}, {Gilmore}, {Randich}, {Asplund},
  {Binney}, {Drew}, {Feltzing}, {Ferguson}, {Jeffries}, {Micela}, {Negueruela},
  {Prusti}, {Rix}, {Alfaro}, {Babusiaux}, {Bensby}, {Blomme}, {Flaccomio},
  {Fran{\c c}ois}, {Irwin}, {Koposov}, {Walton}, {Bayo}, {Carraro}, {Costado},
  {Damiani}, {Edvardsson}, {Hourihane}, {Jackson}, {Lewis}, {Lind}, {Marconi},
  {Martayan}, {Monaco}, {Morbidelli}, {Prisinzano}, \& {Zaggia}}]{smi14}
{Smiljanic}, R., {Korn}, A.~J., {Bergemann}, M., {et~al.} 2014, \aap, 570, A122

\bibitem[{{Steinmetz} {et~al.}(2006){Steinmetz}, {Zwitter}, {Siebert},
  {Watson}, {Freeman}, {Munari}, {Campbell}, {Williams}, {Seabroke}, {Wyse},
  {Parker}, {Bienaym{\'e}}, {Roeser}, {Gibson}, {Gilmore}, {Grebel}, {Helmi},
  {Navarro}, {Burton}, {Cass}, {Dawe}, {Fiegert}, {Hartley}, {Russell},
  {Saunders}, {Enke}, {Bailin}, {Binney}, {Bland-Hawthorn}, {Boeche}, {Dehnen},
  {Eisenstein}, {Evans}, {Fiorucci}, {Fulbright}, {Gerhard}, {Jauregi}, {Kelz},
  {Mijovi{\'c}}, {Minchev}, {Parmentier}, {Pe{\~n}arrubia}, {Quillen}, {Read},
  {Ruchti}, {Scholz}, {Siviero}, {Smith}, {Sordo}, {Veltz}, {Vidrih}, {von
  Berlepsch}, {Boyle}, \& {Schilbach}}]{ste06}
{Steinmetz}, M., {Zwitter}, T., {Siebert}, A., {et~al.} 2006, \aj, 132, 1645

\bibitem[{{Ting} {et~al.}(2015){Ting}, {Conroy}, \& {Goodman}}]{tin15a}
{Ting}, Y.-S., {Conroy}, C., \& {Goodman}, A. 2015, \apj, 807, 104

\bibitem[{{Ting} {et~al.}(2016{\natexlab{a}}){Ting}, {Conroy}, \&
  {Rix}}]{tin16}
{Ting}, Y.-S., {Conroy}, C., \& {Rix}, H.-W. 2016{\natexlab{a}}, \apj, 826, 83

\bibitem[{{Ting} {et~al.}(2016{\natexlab{b}}){Ting}, {Conroy}, \&
  {Rix}}]{tin15b}
---. 2016{\natexlab{b}}, \apj, 816, 10

\bibitem[{{Ting} {et~al.}(2012){Ting}, {Freeman}, {Kobayashi}, {De Silva}, \&
  {Bland-Hawthorn}}]{tin12a}
{Ting}, Y.-S., {Freeman}, K.~C., {Kobayashi}, C., {De Silva}, G.~M., \&
  {Bland-Hawthorn}, J. 2012, \mnras, 421, 1231

\bibitem[{{Venn} {et~al.}(2004){Venn}, {Irwin}, {Shetrone}, {Tout}, {Hill}, \&
  {Tolstoy}}]{ven04}
{Venn}, K.~A., {Irwin}, M., {Shetrone}, M.~D., {et~al.} 2004, \aj, 128, 1177

\bibitem[{{Ventura} {et~al.}(2015){Ventura}, {Karakas}, {Dell'Agli}, {Boyer},
  {Garc{\'{\i}}a-Hern{\'a}ndez}, {Di Criscienzo}, \& {Schneider}}]{ven15}
{Ventura}, P., {Karakas}, A.~I., {Dell'Agli}, F., {et~al.} 2015, \mnras, 450,
  3181

\bibitem[{{Woosley} \& {Weaver}(1995)}]{woo95}
{Woosley}, S.~E., \& {Weaver}, T.~A. 1995, \apjs, 101, 181

\bibitem[{{Yang} {et~al.}(2013){Yang}, {Kirby}, {Guhathakurta}, {Peng}, \&
  {Cheng}}]{yan13}
{Yang}, L., {Kirby}, E.~N., {Guhathakurta}, P., {Peng}, E.~W., \& {Cheng}, L.
  2013, \apj, 768, 4

\bibitem[{{Yanny} {et~al.}(2009){Yanny}, {Rockosi}, {Newberg}, {Knapp},
  {Adelman-McCarthy}, {Alcorn}, {Allam}, {Allende Prieto}, {An}, {Anderson},
  {Anderson}, {Bailer-Jones}, {Bastian}, {Beers}, {Bell}, {Belokurov},
  {Bizyaev}, {Blythe}, {Bochanski}, {Boroski}, {Brinchmann}, {Brinkmann},
  {Brewington}, {Carey}, {Cudworth}, {Evans}, {Evans}, {Gates}, {G{\"a}nsicke},
  {Gillespie}, {Gilmore}, {Nebot Gomez-Moran}, {Grebel}, {Greenwell}, {Gunn},
  {Jordan}, {Jordan}, {Harding}, {Harris}, {Hendry}, {Holder}, {Ivans},
  {Ivezi{\v c}}, {Jester}, {Johnson}, {Kent}, {Kleinman}, {Kniazev},
  {Krzesinski}, {Kron}, {Kuropatkin}, {Lebedeva}, {Lee}, {French Leger},
  {L{\'e}pine}, {Levine}, {Lin}, {Long}, {Loomis}, {Lupton}, {Malanushenko},
  {Malanushenko}, {Margon}, {Martinez-Delgado}, {McGehee}, {Monet}, {Morrison},
  {Munn}, {Neilsen}, {Nitta}, {Norris}, {Oravetz}, {Owen}, {Padmanabhan},
  {Pan}, {Peterson}, {Pier}, {Platson}, {Re Fiorentin}, {Richards}, {Rix},
  {Schlegel}, {Schneider}, {Schreiber}, {Schwope}, {Sibley}, {Simmons},
  {Snedden}, {Allyn Smith}, {Stark}, {Stauffer}, {Steinmetz}, {Stoughton},
  {SubbaRao}, {Szalay}, {Szkody}, {Thakar}, {Sivarani}, {Tucker}, {Uomoto},
  {Vanden Berk}, {Vidrih}, {Wadadekar}, {Watters}, {Wilhelm}, {Wyse}, {Yarger},
  \& {Zucker}}]{yan09}
{Yanny}, B., {Rockosi}, C., {Newberg}, H.~J., {et~al.} 2009, \aj, 137, 4377

\end{thebibliography}

\end{document}